\documentclass[sigconf]{acmart}

\AtBeginDocument{%
  \providecommand\BibTeX{{%
    \normalfont B\kern-0.5em{\scshape i\kern-0.25em b}\kern-0.8em\TeX}}}

\acmConference[KDD '22]{KDD'22}{August 14--18, 2022}{KDD, Washington DC, WA}

\usepackage{enumitem}
\setlist[itemize]{leftmargin=*}
\setlist[enumerate]{leftmargin=*}
\usepackage{multicol}
\usepackage{multirow}
\usepackage{algorithm}
\usepackage{algpseudocode}
\usepackage{makecell}

\begin{document}

\title{MetaV: A Meta-Verifier Approach to Task-Agnostic Model Fingerprinting}

\author{\rm{Xudong Pan, Yifan Yan, Mi Zhang, Min Yang} \\ \textit{School of Computer Science, Fudan University, China} \\ 
\{xdpan18, yanyf20, mi\_zhang, m\_yang\}@fudan.edu.cn 
} 
\renewcommand{\shortauthors}{Pan et al.}

\begin{abstract}
Protecting the intellectual property (IP) of deep neural networks (DNN) becomes an urgent concern for IT corporations. For model piracy forensics, previous model fingerprinting schemes are commonly based on adversarial examples constructed for the owner's model as the \textit{fingerprint}, and verify whether a suspect model is indeed pirated from the original model by matching the behavioral pattern on the fingerprint examples between one another. However, these methods heavily rely on the characteristics of classification tasks which inhibits their application to more general scenarios. To address this issue, we present MetaV, the first task-agnostic model fingerprinting framework which enables fingerprinting on a much wider range of DNNs independent from the downstream learning task, and exhibits strong robustness against a variety of ownership obfuscation techniques. Specifically, we generalize previous schemes into two critical design components in MetaV: the \textit{adaptive fingerprint} and the \textit{meta-verifier}, which are jointly optimized such that the meta-verifier learns to determine whether a suspect model is stolen based on the concatenated outputs of the suspect model on the adaptive fingerprint. As a key of being task-agnostic, the full process makes no assumption on the model internals in the ensemble only if they have the same input and output dimensions. Spanning classification, regression and generative modeling, extensive experimental results validate the substantially improved performance of MetaV over the state-of-the-art fingerprinting schemes and demonstrate the enhanced generality of MetaV for providing task-agnostic fingerprinting. For example, on fingerprinting ResNet-18 trained for skin cancer diagnosis, MetaV achieves simultaneously $100\%$ true positives and $100\%$ true negatives on a diverse test set of $70$ suspect models, achieving an about $220\%$ relative improvement in ARUC in comparison to the optimal baseline.                    
\end{abstract}



\maketitle

\section{Introduction}
In the past decades, deep learning finds a wide application in a variety of mission-critical scenarios in the real world, including autonomous driving \cite{Cao2019AdversarialSA}, finance \cite{Heaton2016DeepLF}, intelligent healthcare \cite{Esteva2017DermatologistlevelCO}, and many more. In the ever-evolving trend of applying deep learning in IT industry, increasingly more high-ended computing power and massive amounts of well-annotated data are devoted to the construction of deep neural networks (DNN) \cite{Real2019RegularizedEF,Devlin2019BERTPO,Zhou2021InformerBE}, which are later deployed as prediction APIs, i.e., Machine-Learning-as-a-Service (MLaaS), to provide intelligent service for profiting. Considering the substantial training costs, many IT corporations as the model owners become aware of the importance of protecting the confidentiality of those well-trained DNN, as an inseparable part of their intelligent property (IP). Threateningly, even with careful access control, an attacker can still pirate the working DNN behind an online intelligent service by conducting system \cite{Yan2020CacheTL,Jeong2021NeuralNS} or algorithmic attacks \cite{Tramr2016StealingML,Yu2020CloudLeakLD}. 

Orthogonal to the advances in protecting DNNs against model privacy \cite{Juuti2019PRADAPA}, \textit{model watermarking} and \textit{model fingerprinting} are two fast-developing techniques for model piracy forensics. Applying model watermarking, the model owner embeds a secret into his/her owned model (i.e., the \textit{target model}). Once the ownership of a DNN model is in doubt (i.e., the \textit{suspect model}), a trusted third party verifies the existence of the exclusively known secret in the suspect model to determine the actual ownership. From \citet{Uchida2017EmbeddingWI}, previous works devise different types of secrets (e.g., a specific function or a specific parameter pattern) into various parts of a DNN, which we briefly survey in Section \ref{sec:related}. However, because model watermarking unavoidably modifies the original parameters of a well-trained DNN for secret embedding, the otherwise optimal accuracy would be slightly degraded, causing an unacceptable trade-off for mission-critical tasks in healthcare and traffic \cite{Cao2019IPGuardPT}.     

Complemental to model watermarking, model fingerprinting is a passive forensic technique against model piracy, which in general tests whether certain \textit{fingerprint} of the target model are present in a suspect model, which would help collect essential evidence of model piracy in the wild before filing a lawsuit. As a key difference from model watermarking, the fingerprint is innate but not embedded to the target model. In other words, no modifications on the target model are conducted during fingerprinting, which provably preserves the normal utility of well-trained DNNs. Thanks to this desirable characteristic, model fingerprinting arises as a booming direction in model protection from the last year, which attracts increasing research efforts from different backgrounds \cite{Cao2019IPGuardPT,Li2021ModelDiffTD,Wang2021FingerprintingDN,Lukas2019DeepNN}. 

Following the fingerprinting framework in \citet{Cao2019IPGuardPT}, previous schemes mostly focus on fingerprinting classifiers by constructing a special set of \textit{adversarial examples} \cite{Szegedy2014IntriguingPO}, i.e., normal examples added with human-imperceptible perturbations which cause misclassification of the target classifier, as the fingerprint, and verifying whether a suspect model is indeed stolen from the original model by matching the behavioral pattern, e.g., the predicted labels \cite{Lukas2019DeepNN,Wang2021FingerprintingDN} or probability vector similarity \cite{Li2021ModelDiffTD}, on the fingerprint examples. Despite their pioneering contributions to model IP protection, \emph{existing schemes are however limited to fingerprinting DNNs for other important downstream tasks except for classification}, mainly because they commonly rely on concepts like adversarial examples and classification boundary which have no direct counterparts in other typical learning tasks such as regression and generative modeling. With recent years witnessing the fast trend of distribution, deployment and redistribution of DNNs in nowadays deep learning ecosystem, how to conduct forensics on the improper reuse and illegal piracy for a more general set of DNNs poses an urgent open challenge to address.

\subsection{Our Work} In this paper, we present a meta-verifier approach to task-agnostic model fingerprinting scheme (dubbed as \textit{MetaV}), which for the first time enables fingerprinting on a much wider range of DNNs independent from the downstream learning task, and by design implements the robustness against a variety of ownership obfuscation techniques possibly adopted by the adversary.

To realize task-agnostic model fingerprinting, we generalize the idea of using adversarial examples and the corresponding classification results for fingerprinting respectively into two critical design components in MetaV, i.e., the \textit{adaptive fingerprint} and the \textit{meta-verifier}. Concisely, the adaptive fingerprint is a set of trainable inputs to the suspect model, the concatenated outputs of the suspect model on which are classified by the meta-verifier to be \textit{True} or \textit{False}, where \textit{True} implies the suspect model is indeed stolen (i.e., \textit{positive suspect model}), and \textit{False} implies the suspect model is independent from the original model (i.e., \textit{negative suspect model}). 

To implement the design principles above, the adaptive fingerprint and the meta-verifier are jointly optimized on an ensemble composed of the target model, the positive and the negative suspect models which are virtually generated by MetaV during the fingerprinting construction phase. Specifically, the positive suspect models in the ensemble are crafted by post-processing the target model with a number of popular ownership obfuscation techniques, e.g., compression \cite{Han2015LearningBW,Li2017PruningFF}, fine-tuning, partial retraining and distillation \cite{Hinton2015DistillingTK}, while the irrelevant models are independently trained from scratch on similar learning tasks to the target model. As the full construction process of MetaV makes no assumption on the model internals or functions in the ensemble only if they have the same input and output dimensions, MetaV is therefore applicable independent of the downstream tasks for which the DNN is designed. Moreover, by permitting more types of obfuscation techniques in producing the stolen models for the model ensemble, our proposed MetaV is by construction robust against a diverse set of existing obfuscation techniques, with the potential to evolve along with future adversarial techniques. 

In summary, we mainly make the following contributions: 
\begin{itemize}
    \item We present MetaV, the first task-agnostic fingerprinting framework with adaptive robustness against a variety of ownership obfuscation techniques, to substantially advance the cutting-edge model fingerprinting capability to a much broader set of DNNs for arbitrary downstream tasks.
    \item We generalize the existing fingerprinting schemes based on adversarial examples to a more universal fingerprinting framework based on the adaptive fingerprint and the meta-verifier, which are jointly optimized on an ensemble of the target model, the positive and negative suspect models crafted by the model owner to serve as a highly effective and robust fingerprint for the target model.    
    \item We extensively evaluate the performance of MetaV on practical scenarios spanning classification, regression and generative modeling. Besides the unique contribution of MetaV in providing task-agnostic fingerprinting, MetaV brings noticeable improvement over all the state-of-the-art fingerprinting schemes on DNN classifiers. For example, on fingerprinting ResNet-18 \cite{He2016DeepRL} trained for skin cancer diagnosis \cite{Esteva2017DermatologistlevelCO}, MetaV achieves simultaneously $100\%$ true positives and $100\%$ true negatives on a diverse test set of $70$ suspect models, with an about $220\%$ relative improvement in ARUC compared to the optimal baseline.

\end{itemize}



\section{Related Works}
\label{sec:related}
\subsection{Model Fingerprinting}
Recently, a number of fingerprinting schemes, mainly based on constructing different types of adversarial examples as the model fingerprint, were proposed to protect the intellectual property of DNN classifiers.
For example, \citet{Cao2019IPGuardPT} propose IPGuard, one of the earliest fingerprinting schemes, to find adversarial examples near the decision boundary of the target classifier. The key assumption of IPGuard is that the target DNN classifier can be uniquely represented by its decision boundary, which is more similar with the classification boundary of a positive suspect model than a negative one.
Different from IPGuard, \citet{Lukas2019DeepNN} and \citet{Zhao2020AFAAF} independently propose to extract so-called \textit{conferrable} adversarial examples from the ensemble of the target classifier and a set of locally trained suspect classifiers. These conferrable adversarial examples, which transfer much better to the positive suspect models than to negative ones, can be regarded as a unique link between the target model and the positive suspect models. Besides, \citet{Wang2021FingerprintingDN} utilize the geometry characteristics inherited in the DeepFool algorithm to construct adversarial examples as the fingerprint \cite{Wang2021FingerprintingDN}, while \citet{Li2021ModelDiffTD} leverage the similarity between models in terms of the probability vectors on test inputs for piracy detection \cite{Li2021ModelDiffTD}. More detailed surveys can be found in \cite{Boenisch2020ASO,Regazzoni2021ProtectingAI}. In this work, our proposed MetaV generalizes the aforementioned fingerprinting techniques to abstract the usage of adversarial examples and the classification results into the adaptive fingerprint and the trainable meta-verifiers, which is applicable to arbitrary DNN models in a task-agnostic way.

\subsection{Model Watermarking}
Orthogonal to model fingerprinting, model watermarking embeds a watermark into the trained model before it is released, which potentially sacrifices the utility of the model. Previous works on model watermarking explore various types of watermarks such as secret bit strings \cite{Uchida2017EmbeddingWI,Rouhani2018DeepSignsAG}, generated serial numbers \cite{Xu2020IdentityBF} and unrelated or slightly modified sample sets \cite{Adi2018TurningYW,Zhang2018BackdoorWatermark}. These identifying codes are then encoded secretly into the least significant bit of the weight \cite{Uchida2017EmbeddingWI}, the distribution of outputs at the intermediate \cite{Rouhani2018DeepSignsAG} or the full layers \cite{Adi2018TurningYW,Zhang2018BackdoorWatermark,Xu2020IdentityBF}
\section{Security Settings}
\label{sec:prelim}
\subsection{Backgrounds \& Notions} Model fingerprinting is a multi-party security game among a \textit{model owner}, a \textit{verifier} and an \textit{attacker}. Initially, a model owner devotes computing power and well-curated training data to build its own DNN $F:\mathcal{X}\to\mathcal{Y}$ for a certain downstream task, crowning the obtained model $F$ as an inseparable part of the model owner's IP. Following the nomenclature in \citet{Cao2019IPGuardPT}, we refer to the model $F$ as the \textit{target model}. In nowadays deep learning ecosystem, the model owner can deploy the target model at a third-party platform like Amazon AWS as a prediction API to gain monetary profits. However, the profits may also serve as incentives on the potential attacker to conduct model piracy via, e.g., software/hardware vulnerabilities \cite{Yan2020CacheTL,Jeong2021NeuralNS}, social engineering and algorithmic attacks \cite{Tramr2016StealingML,Yu2020CloudLeakLD}. This essentially infringes the IP of the model owner. 

As a rescue, the model owner can delegate a \textit{verifier}, usually played by a trusted third party or the model owner him/herself, to provide model fingerprinting service for model piracy forensics. In general, model fingerprinting determines whether a suspect model $\tilde{F}$ is pirated from the target model $F$ in the two stages:

\begin{itemize}
    \item \textbf{Fingerprint Construction.} At the construction stage, a certain type of model fingerprint encoding the essential characteristics of the target model $F$ is constructed.
    \item \textbf{Fingerprint Verification.} At the verification stage, the suspect model is attested via the prediction API (i.e., black-box access) to determine whether and with what confidence (i.e., \textit{matching rate}) the fingerprint is also present in the suspect model $\tilde{F}$.   
\end{itemize}


\begin{figure*}[t]
\begin{center}
\includegraphics[width=1.0\textwidth]{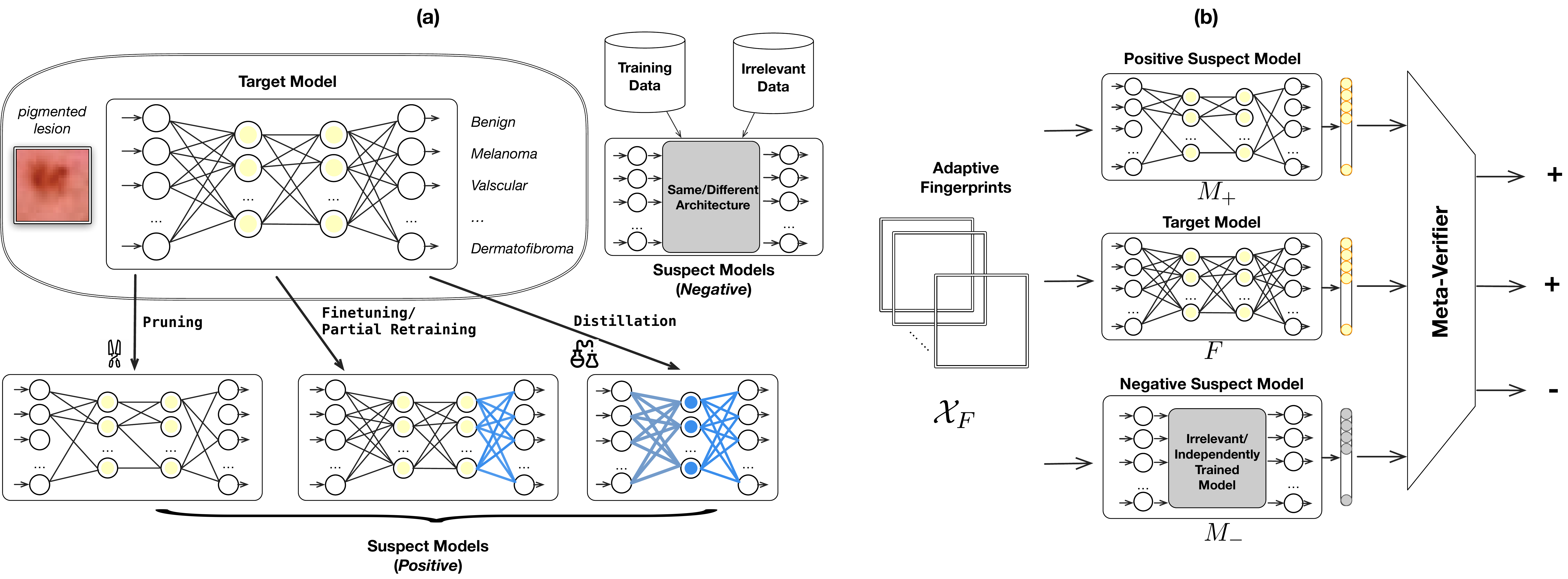}
\caption{The general pipeline of our proposed MetaV: \textbf{(a)} model ensemble preparation and  \textbf{(b)} fingerprint construction.}
\label{fig:utaf_overview}
\end{center}
\end{figure*}

\subsection{Threat Model} We mainly consider the following threat model in this paper.
\begin{itemize}
    \item \textbf{Attacker's Capability.} We assume the attacker would apply a variety of model post-processing techniques to obfuscate the ownership of the stolen model (detailed in the subsequent part) after he/she successfully steals the target model from an online prediction API. Such an obfuscated model is called a \textit{positive suspect model}. Correspondingly, a suspect model independently trained by another honest model owner is called a \textit{negative suspect model}. The attacker is assumed to answer any queries to his/her provided prediction API, as more queries served by the API bring more monetary profits.  
    \item \textbf{Verifier's Capability.} Following, e.g., \citet{Cao2019IPGuardPT}, we assume the verifier has a white-box access to the target model while a black-box access to the suspect models via their prediction APIs. To relax their assumptions, we \textit{do not} assume the internal architecture or the type of downstream tasks of the target model.   
\end{itemize}

\subsection{Adversarial Techniques} Integrating existing ownership obfuscation techniques studied in previous works, we mainly cover the following classes of adversarial techniques which the attacker is likely to adopt. 
\begin{itemize}
    \item \textbf{Model Compression}: Compression-based obfuscation adopts weight (filter) pruning \cite{Li2017PruningFF} to remove a certain ratio of small weights (filters) in a DNN, which would largely preserve the utility of the obfuscated model on the learning task \cite{Han2015LearningBW} 
    and inhibits heuristic-based fingerprinting from verifying the model owernship based on parameter comparison \cite{Cao2019IPGuardPT}.   
    \item \textbf{Fine-Tuning \& Partial Retraining}: To obfuscate the behavioral pattern of a DNN, attackers may resume the training of the stolen model on public data collected from a similar domain of the training data. Specifically, to fine-tune the last $K$ layers of a trained DNN, the parameters of the last $K$ layers are further updated according to the learning objective with the other layers fixed. In comparison, during the partial retraining, the parameters of the last $K$ layers are first randomly initialized before the training is resumed. Due to the non-convexity of deep learning \cite{Choromaska2015TheLS}, both the fine-tuned and partially-retrained models may fall into a different local optimum, preserve the original utility, but exhibit divergent prediction behaviors \cite{Wang2021FingerprintingDN}. 
    \item \textbf{Model Distillation}: Distillation-based obfuscation adopts the knowledge distillation strategies \cite{Hinton2015DistillingTK,Gou2021KnowledgeDA} by viewing the stolen model or the corresponding prediction API as the \textit{teacher} model, and a DNN of a different architecture as the \textit{student} model \cite{Li2021ModelDiffTD}. Via distillation, the learned knowledge in the target model is inherited by the student model, which 
   exacerbates the obfuscation of model ownership due to the transformed model architecture and the correspondingly altered predictive behaviors \cite{Lukas2019DeepNN}.  
\end{itemize}

\section{Our Methodology for Task-Agnostic Fingerprinting}
\label{sec:method}

\subsection{Overview of MetaV} As a generalization of previous fingerprinting schemes, our proposed MetaV abstracts the usage of adversarial examples and the corresponding prediction results as the fingerprint into two critical components respectively: (i) \textit{adaptive fingerprint}, i.e., a set of trainable inputs to the target/suspect model(s), which we denote as $\mathcal{X}_{F} := (x_F^{1}, \hdots, x_F^{N})$ with $x_{F}^{i} \in \mathcal{X}$ and $N$ called the number of fingerprint examples, and (ii) \textit{meta-verifier}, i.e., a binary classifier which takes the concatenated outputs of a suspect model on the adaptive fingerprint as its input and predicts whether the suspect model is positive or negative, denoted as $\mathcal{V}:\mathcal{Y}^{n} \to \mathbb{S}^{2}$, a $2$-dimensional probability simplex $\{(p_{-}, p_{+})| p_{-} + p_{+} = 1, 0 \le p_{+},p_{-} \le 1\}$. As illustrated in Fig. \ref{fig:utaf_overview}, the general pipeline of MetaV mainly consists of the following three key stages:
\begin{itemize}
    \item \textbf{Stage 1.} \textit{(Model Ensemble Preparation)} As Fig. \ref{fig:utaf_overview}(a) shows, we first craft a number of positive and negative suspect models from the target model with the aid of public data from the same domain of the owner's training data. At the end of this stage, we obtain the sets of positive and negative suspect models, i.e., $\mathcal{M}_{+}$ and $\mathcal{M}_{-}$.   
    \item \textbf{Stage 2.}  \textit{(Fingerprint Construction in MetaV)} As Fig. \ref{fig:utaf_overview}(b) shows, we then jointly optimize the adaptive fingerprint $\mathcal{X}_F$ and the meta-verifier $\mathcal{V}$ to satisfy: For both the target model or any suspect model in $\mathcal{M}_{+}$, the meta-verifier $V$ is trained to predict \textit{True}, i.e., $p_{+} > p_{-}$, on the concatenated outputs of the model on examples in $\mathcal{X}_{F}$, and vice versa for any suspect model in $\mathcal{M}_{-}$. At the end of this stage, we obtain optimized adaptive fingerprint and the corresponding meta-verifier, i.e., $\mathcal{X}_{F}^{*}$ and $\mathcal{V}^{*}$ respectively. We call $(\mathcal{X}_{F}^{*}, \mathcal{V}^{*})$ a \textit{fingerprinting pair}.     
    \item \textbf{Stage 3.} \textit{(Fingerprint Verification in MetaV)} Finally, with the optimized fingerprinting pair $(\mathcal{X}_{F}^{*}, \mathcal{V}^{*})$, we verify whether and with what matching rate a suspect model $\tilde{F}$ is a stolen version or an independently trained one by querying the prediction API with the fingerprint examples in $\mathcal{X}_{F}$. The received prediction results are then concatenated and input to the meta-verifier to predict $(\tilde{p}_{-}, \tilde{p}_{+}) = \mathcal{V}(\tilde{F}(x_F^{1}) \oplus \hdots \oplus \tilde{F}(x_F^{N}))$. When $\tilde{p}_{+}$ is larger than a predefined threshold $\rho$, the verification process outputs \textit{True} to claim possible model piracy behind the tested prediction API, or other the process outputs \textit{False} to assert the fidelity of the suspect model.    
\end{itemize}
In the following sections, we elaborate on the detailed methodology for the first two stages of MetaV. 

\subsection{Model Ensemble Preparation}
To construct the adaptive fingerprint and the meta-verifier with simultaneously high robustness and uniqueness, MetaV is first required to collaborate with the model owner to prepare a diverse set of positive and negative suspect models. Intuitively, with a more representative set of positive suspect models, the obtained fingerprinting pair would stay robust against a wider range of ownership obfuscation techniques, resulting in higher true positives. Alternatively, more representative negative suspect models would lower the probability of the learned fingerprinting pair to be present in other irrelevant models, which therefore reduces the true negatives of MetaV. Specifically, the positive and negative suspect models are constructed as follows. 
\begin{itemize}
    \item \textbf{Prepare Positive Suspect Models}. We derive a representative set of positive suspect models by randomly applying one or more common ownership obfuscation techniques mentioned in Section \ref{sec:prelim} to the target model $F$. The applied obfuscation techniques are recommended to cover a wider range of hyperparameter configurations for better robustness. For example, we apply weight and filter pruning to the target model with different pruning ratios. As a notation, we denote the full set of obfuscation techniques for preparing the positive suspect models as $\mathcal{T}$. Therefore, we have $\mathcal{M}_{+} := \{T\circ{F} | T \in \mathcal{T}\}$. Section \ref{sec:eval_setting} presents the detailed composition of $\mathcal{T}$ in the evaluation settings part. 
    \item \textbf{Prepare Negative Suspect Models}. We recommend three complemental sources to collect a diverse set of negative suspect models for the fingerprint construction stage of MetaV. First, MetaV may request the model owner to train a moderate number of relatively small-scale DNNs on the same training dataset of the target model. For IP protection of the target model, it would be reasonable for the model owner to devote additional computing power to collaborate with a trusted third-party verifier. Second, MetaV can also download a number of pretrained models from online sources (e.g., PyTorch Hub) and fine-tune these models on the domain-relevant public data to serve as the negative suspect models. Moreover, MetaV may consider incorporate a proportion of irrelevant publicly available models into the set of negative suspect models to further enhance the uniqueness of the obtained fingerprinting pair. We denote the prepared negative suspect models as $\mathcal{M}_{-}$. 
\end{itemize}

\begin{algorithm}[t]
\begin{algorithmic}[1]
\State{\textbf{Input:} A prepared model ensemble $\mathcal{M}_{-} \cup \{F\} \cup \mathcal{M}_{+}$, the number of adaptive fingerprints $N$, the input/output dimension of the models $d_\text{in}, d_\text{out}$, the number of iterations $L$ and the learning rate $\lambda$.}
\State{\textbf{Output:} The optimal fingerprinting pair $(\mathcal{X}_F^{*}, \mathcal{V}^{*})$.}
\State{Initialize real-valued variables $W_0 = (w_i)_{i=1}^{N}$.}
\State{Initialize a meta-verifier $\mathcal{V}(\cdot; \Theta_0):\mathcal{Y}^{N\times{d}_\text{out}}\to \mathbb{S}^{2}$ with parameters $\Theta_0$} \Comment{We implement $\mathcal{V}$ as a fully-connected neural network with a softmax output.}
\State{Initialize Adam optimizers $\text{Opt}_{W}, \text{Opt}_{\Theta}$ of a learning rate $\lambda$.}

\For{$t$ in $\{0, \hdots, L-1\}$}
\State{Sample a tuple of models $(M_{-}, F, M_{+})$ from $\mathcal{M}_{-}$, $\{F\}$ and $\mathcal{M}_{+}$ respectively.}
\State{For each $i \in [N]$, calculate $x_{F}^{i} = \tanh(w_i)$.}
\State{$\ell(W_t, \Theta_t) = \log{p_{+}(M_{+})} + \log{p_{+}(F)} + \log{p_{-}(M_{-})}$} \Comment{ $(p_{-}(M), p_{+}(M)) = \mathcal{V}(M(x_F^{1}) \oplus \hdots \oplus M(x_F^{N}))$}
\State{$W_{t+1} \gets \text{Opt}_{W}(\ell, W_t)$.}
\State{$\Theta_{t+1} \gets \text{Opt}_{\Theta}(\ell, \Theta_t)$.}
\EndFor

\State{For each $i \in [N]$, $x_{F}^{i,*} = \tanh(w_i^{L})$.}
\State{$\Theta^{*} \gets {\Theta^{L}}$.}
\State{\textbf{Return: }($\mathcal{X}_F^{*}, \mathcal{V}(\cdot; \Theta^{*})$)}
\caption{The algorithmic details of MetaV's fingerprint construction stage.}
\label{alg:utaf_construction}
\end{algorithmic}
\end{algorithm}


\subsection{Fingerprint Construction in MetaV} In this part, we detail the learning objective and the optimization algorithm for training the fingerprinting pair $(\mathcal{X}_F, \mathcal{V})$ on the model ensemble $\mathcal{M}_{-} \cup \{F\} \cup {\mathcal{M}_{+}}$ prepared in the first stage. As Fig. \ref{fig:utaf_overview}(b) shows, the learning objective of MetaV is viewed as a binary classification problem. To formulate, we introduce an additional label $s$ for each model, which takes value in  $\{+, -\}$ (literally, \textit{positive} and \textit{negative} respectively). Specifically, we label an arbitrary model $M_{+} \in \mathcal{M}_{+} \cup \{F\}$ as $+$ and an arbitrary model $M_{-} \in \mathcal{M}_{-}$ as $-$. To supervise the adaptive fingerprint and the meta-verifier with the labels, we solve the learning objective:
\begin{align}
     \text{arg}\max_{\mathcal{X}_F, \mathcal{V}} \log{p_{+}(F)} +  \frac{1}{|\mathcal{M}_{+}|} \sum_{M_{+}\in\mathcal{M}_{+}}\log{p_{+}(M_{+})} \nonumber \\ +  \frac{1}{|\mathcal{M}_{-}|} \sum_{M_{-}\in\mathcal{M}_{-}}\log{p_{-}(M_{-})},
     \label{eq:full_loss}
\end{align}
where $(p_{-}(M), p_{+}(M)) = \mathcal{V}(M(x_F^{1}) \oplus \hdots \oplus M(x_F^{N}))$, the prediction from the meta-verifier on the concatenated outputs of a model under test on the adaptive fingerprint. Intuitively, the learning objective above encourages the meta-verifier to output a $p_{+}$ higher than $p_{-}$ when the model under test is a positive suspect model or the target model, and vice versa for a negative suspect model.

As the learning objective above is fully derivative w.r.t. the adaptive fingerprint $\mathcal{X}_{F}$ and the parameters of the meta-verifier, we leverage off-the-shelf non-convex optimizers (e.g., Adam \cite{Kingma2015AdamAM}) for gradient-based optimization. However, we notice it is resource-consuming to conduct back-propagation over the whole model ensembles in each optimization step. As an alternative, we reformulate the batched learning objective in (\ref{eq:full_loss}) as a stochastic objective with randomness in a tuple of $(M_{-}, F, M_{+})$ uniformly sampled from $\mathcal{M}_{-}$, $\{F\}$ and $\mathcal{M}_{+}$ in each iteration. Besides, we adopt the reparametrization trick in \citet{Carlini2017TowardsET} to constrain the adaptive fingerprint in the problem space $\mathcal{X}$.  Taking $\mathcal{X} := [-1, 1]^{d_{\text{in}}}$, a common case in computer vision for example, Algorithm \ref{alg:utaf_construction} presents the details of the optimization algorithm.


\section{Evaluation Settings}
\label{sec:eval_setting}
\subsection{Scenarios and Datasets} Table \ref{tab:scenarios} provides an overview on the three scenarios, i.e., \textit{skin cancer diagnosis} (\citet{Yang2020MedMNISTCD},classification), \textit{warfarin dose prediction} (\citet{WhirlCarrillo2012PharmacogenomicsKF}, regression), and \textit{fashion generation} (\citet{Xiao2017FashionMNISTAN}, generative modeling), covered in the evaluation sections. The information of the datasets are concisely introduced below.
\begin{itemize}
    \item \textbf{Skin Cancer Diagnosis} (\textit{abbrev.} \textbf{Skin}). The first scenario covers the usage of deep convolutional neural network (CNN) for skin cancer diagnosis. According to \cite{Yang2020MedMNISTCD}, we train a ResNet-18 \cite{He2016DeepRL} as the target model on DermaMNIST \cite{Yang2020MedMNISTCD}, which consists of $10005$ multi-source dermatoscopic images of common pigmented skin lesions imaging dataset. The input size is originally $3\times{28}\times{28}$, which is upsampled to be $3\times{224}\times{224}$ to fit the input shape of a standard ResNet-18 architecture implemented in torchvision \cite{pytorch_hub}. The task is a $7$-class classification task.
    \item \textbf{Warfarin Dose Prediction} (\textit{abbrev.} \textbf{Warfarin}). The second scenario covers the usage of FCN for warfarin dose prediction, which is a safety-critical regression task that helps predict the proper individualised warfarin dosing according to the demographic and physiological record of the patients (e.g., weight, age and genetics). We use the International Warfarin Pharmacogenetics Consortium (IWPC) dataset \cite{WhirlCarrillo2012PharmacogenomicsKF}, which is a public dataset composed of $31$-dimensional features of $6256$ patients and is widely used for researches in automated warfarin dosing. According to \citet{Truda2019WarfarinDE}, we use a three-layer multi-layer perception (MLP) with ReLU as the target model, with its hidden layer composed of $100$ neurons. As a notation, we denote the architecture as $(31$-$100$-$1)$. The target model learns to predict the value of proper warfarin dosing, which is a non-negative real-valued scalar with its value in $(0, 300.0]$.
    \item \textbf{Fashion Generation} (\textit{abbrev.} \textbf{Fashion}) The final scenario covers the usage of FCN for generative modeling. We choose \cite{Xiao2017FashionMNISTAN}, which consists of $60000$  images for articles of clothing of size $28\times{28}$. We train a DCGAN-like architecture \cite{Radford2016UnsupervisedRL} for generative modeling on this task. We solely view the generator as the target model, as a well-trained generator represents more the IP of the model owner because it can be directly used to generate realistic images without the aid of the discriminator. The detailed DCGAN architecture we use is demonstrated in Table \ref{tab:app:dcgan}.
\end{itemize}

\begin{table}[htbp]
  \centering
  \caption{The detailed architecture of DCGAN on Fashion, which is described by convention of PyTorch.}
  \scalebox{0.85}{
    \begin{tabular}{cl}
    \toprule
    \multirow{11}[2]{*}{\textbf{Generator}} & nn.ConvTranspose2d(100, 128, 4, 1, 0, bias=False) \\
          & nn.BatchNorm2d(128) \\
          & nn.ReLU() \\
          & nn.ConvTranspose2d(128, 64, 3, 2, 1, bias=False) \\
          & nn.BatchNorm2d(64) \\
          & nn.ReLU() \\
          & nn.ConvTranspose2d(64, 32, 4, 2, 1, bias=False) \\
          & nn.BatchNorm2d(32) \\
          & nn.ReLU() \\
          & nn.ConvTranspose2d(32, 1, 4, 2, 1, bias=False) \\
          & nn.Tanh() \\
    \midrule
    \multirow{10}[1]{*}{\textbf{Discriminator}} & nn.Conv2d(1, 32, 4, 2, 1, bias=False) \\
          & nn.LeakyReLU(0.2) \\
          & nn.Conv2d(32, 64, 4, 2, 1, bias=False) \\
          & nn.BatchNorm2d(64) \\
          & nn.LeakyReLU(0.2) \\
          & nn.Conv2d(64, 128, 3, 2, 1, bias=False) \\
          & nn.BatchNorm2d(128) \\
          & nn.LeakyReLU(0.2) \\
          & nn.Conv2d(128, 1, 4, 1, 0, bias=False) \\
          & nn.Sigmoid() \\
    \bottomrule
    \end{tabular}}%
  \label{tab:app:dcgan}%
\end{table}%


\begin{table}[ht]
\caption{Learning tasks and datasets in the evaluation.}
\begin{center}
\renewcommand{\arraystretch}{0.5} 
\small
\scalebox{0.9}{
\begin{tabular}{llll}
\toprule
   \textbf{Identifier} & \textbf{Type} &  \textbf{Dataset} &  \textbf{Target Model}   \\ \midrule
   \textbf{Skin}  & Classification& DermaMNIST  & ResNet-18   \\ \midrule
  \textbf{Warfarin}  & Regression  & IWPC Dataset & MLP \\ \midrule
   \textbf{Fashion} & Generative Modeling & FashionMNIST &  DCGAN \\ 
\bottomrule
\end{tabular}}
\end{center}
\label{tab:scenarios}
\end{table}

\subsection{Fingerprinting Benchmarks} For each scenario, we construct a model benchmark composed of $140$ positive/negative suspect models. We split the benchmark randomly by a ratio of $1: 1$ into two independent sets of suspect models for training and testing.  

\noindent$\bullet$ \textbf{Constructing Positive Suspect Models.} Following \citet{Cao2019IPGuardPT} and \citet{Lukas2019DeepNN}, we apply a number of popular ownership obfuscation techniques with a variety of hyperparameter configurations on the target model to derive the positive suspect models:
\begin{enumerate}
    \item \textbf{Compression}: For weight pruning, we vary the ratio of pruned weights from $0.1$ to $0.9$ with a stride of $0.1$. For filter pruning, we choose the ratio of pruned filters from $1/16$ to $15/16$ with a stride of $1/16$.
    \item \textbf{Fine-Tuning \& Partial Retraining}: We consider $4$ types of obfuscation in this category, i.e., fine-tuning/retraining the last layer and fine-tuning/retraining all layers. For both types of retraining, the last one layer is first reset as a randomly initialized layer, after which the model is finetuned according to the configuration. We set the number of epochs for fine-tuning and retraining both as $10$.
    \item \textbf{Distillation}: For each target model, we select $3$-$5$ diverse models with different architectures as the student model. For the ResNet-18 classifier, we follow the classical distillation algorithm in \citet{Hinton2015DistillingTK} to prepare the student model. We do not consider other model distillation algorithms because most of them require the access to the internals of the target model (i.e., the teacher) for distillation, implausible for an attacker who pirates the model from the prediction API. For the multi-layer perception (MLP) as the regressor and the DCGAN \cite{Radford2016UnsupervisedRL} as the generator, we implement the distillation algorithms in \citet{Clark2019BAMBM} and \citet{Aguinaldo2019CompressingGU} respectively. For fine-tuning, partial retraining and distillation, we mutate the random seeds to produce multiple suspect models belonging to the corresponding category.  

\end{enumerate}

\noindent$\bullet$ \textbf{Constructing Negative Suspect Models.} To construct the negative suspect models, we use different random seeds to initialize models of different architectures. We then train the models from scratch respectively on the original training data, on the public data from a similar domain of the training set, and on other irrelevant dataset to obtain a diverse benchmark of negative suspect models. Table \ref{tab:app:model_stat} lists the composition of the suspect models for all the three scenarios. For convenience, we use the following abbreviation: fine-tuning the last layer (=\textit{FTLL}), fine-tuning all layers (=\textit{FTAL}), retraining the last layer (=\textit{RTAL}), retraining all layers (=\textit{RTAL}), weight-pruning (=\textit{WP}), filter-pruning (=\textit{FP}). For constructing distillation-based positive suspect models and independently trained negative suspect models, we implement $3$-$5$ models of diverse architectures and incremental sizes for each of the three target models. For convenience, we index these models as \textit{S, M, L, XL, XLL}. Specifically, these models are:
\begin{itemize}
    \item \textbf{Skin}: \textit{S}=SqueezeNet-1-0 \cite{Iandola2016SqueezeNetAA}; \textit{M}=ResNet-18 (\citet{He2016DeepRL}, the same as the target model); \textit{L}=DenseNet-161 \cite{Huang2017DenselyCC}; \textit{XL}=AlexNet \cite{Krizhevsky2014OneWT}; \textit{XXL}=VGG-16 \cite{Simonyan2015VeryDC}.
    \item \textbf{Warfarin}: \textit{S}=$(31$-$100$-$1)$ (the same as the target model); \textit{M}=$(31$-$100$-$100$-$1)$; \textit{L}=$(31$-$100$-$100$-$100$-$1)$.
    \item \textbf{Fashion}: \textit{S}=Architecture in Table \ref{tab:app:fcn_gan} with $s=1$; \textit{M}=with $s=2$; \textit{L}=with $s=3$; \textit{XL}=the same as the target model in Table \ref{tab:app:dcgan}. 
\end{itemize}

\begin{table}[t]
  \centering
  \caption{Composition of suspect models for each scenario.}
  \scalebox{0.75}{
    \begin{tabular}{clcccc}
    \toprule
          &       &       & \multicolumn{1}{c}{\textbf{Skin}} & \multicolumn{1}{c}{\textbf{Warfarin}} & \multicolumn{1}{c}{\textbf{Fashion}} \\
    \midrule
    \multicolumn{1}{c}{\multirow{11}[10]{*}{\makecell{\textbf{Positive} \\  \textbf{Suspect} \\  \textbf{Models}}}} & \multirow{2}[2]{*}{Fine-tuning} & \multicolumn{1}{l}{FTLL} & $2$     & $2$     & $2$ \\
          &       & \multicolumn{1}{l}{FTAL} & $2$     & $2$     & $2$ \\
\cmidrule{2-6}          & \multirow{2}[2]{*}{Partial Retraining} & \multicolumn{1}{l}{RTLL} & $2$     & $2$     & $2$ \\
          &       & \multicolumn{1}{l}{RTAL} & $2$     & $2$     & $2$ \\
\cmidrule{2-6}          & WP    & \multicolumn{1}{l}{$0.1, 0.2, \hdots, 0.9$} & $9\times{2}$ & $9\times{4}$ & $9\times{2}$ \\
\cmidrule{2-6}          & FP    & \multicolumn{1}{l}{$1/16, \hdots, 15/16$} & $15\times{2}$ & N/A   & $15\times{2}$ \\
\cmidrule{2-6}          & \multirow{5}[2]{*}{Distillation} & \multicolumn{1}{l}{S} & $4$     & $10$    & $6$ \\
          &       & \multicolumn{1}{l}{M} & $4$     & $10$    & $6$ \\
          &       & \multicolumn{1}{l}{L} & $2$     & $6$     & $2$ \\
          &       & \multicolumn{1}{l}{XL} & $2$     & N/A   & N/A \\
          &       & \multicolumn{1}{l}{XXL} & $2$     & N/A   & N/A \\
    \midrule
    \multicolumn{1}{l}{\multirow{6}[2]{*}{{\makecell{\textbf{Negative} \\  \textbf{Suspect} \\  \textbf{Models}}}}} & \multicolumn{1}{l}{\multirow{5}[1]{*}{\makecell{Independently \\ Trained \\ Models}}} & \multicolumn{1}{l}{S} & $20$    & $20$    & $20$ \\
          &       & \multicolumn{1}{l}{M} & $20$    & $20$    & $10$ \\
          &       & \multicolumn{1}{l}{L} & $12$    & $20$    & $10$ \\
          &       & \multicolumn{1}{l}{XL} & $4$     & N/A   & $20$ \\
          &       & \multicolumn{1}{l}{XXL} & $4$     & N/A   & N/A \\
          \cmidrule{2-6} 
          & \multicolumn{2}{l}{Irrelevant Models} & $10$    & $10$    & $10$ \\
    \bottomrule
    \end{tabular}}%
  \label{tab:app:model_stat}%
\end{table}%


\begin{table}[htbp]
  \centering
  \caption{The detailed architecture of the student models for DCGAN on Fashion, which is described by convention of PyTorch ($k=2^{s}$).}
\scalebox{1.0}{  
    \begin{tabular}{cl}
    \toprule
    \multirow{8}[2]{*}{\textbf{Generator}} & nn.Linear(100, 128) \\
          & nn.ReLU() \\
          & nn.Linear($64k$, $128k$) \\
          & nn.ReLU() \\
          & nn.Linear($128k$, $256k$) \\
          & nn.ReLU() \\
          & nn.Linear($256k$, 28x28) \\
          & nn.Sigmoid() \\
    \midrule
    \multirow{7}[2]{*}{\textbf{Discriminator}} & nn.Linear(28x28, $256k$) \\
          & nn.ReLU() \\
          & nn.Linear($256k$, $128k$) \\
          & nn.ReLU() \\
          & nn.Linear($128k$, $64k$) \\
          & nn.ReLU() \\
          & nn.Linear($64k$, $1$) \\
    \bottomrule
    \end{tabular}}%
  \label{tab:app:fcn_gan}%
\end{table}%

\begin{figure*}[t]
\begin{center}
\includegraphics[width=1.0\textwidth]{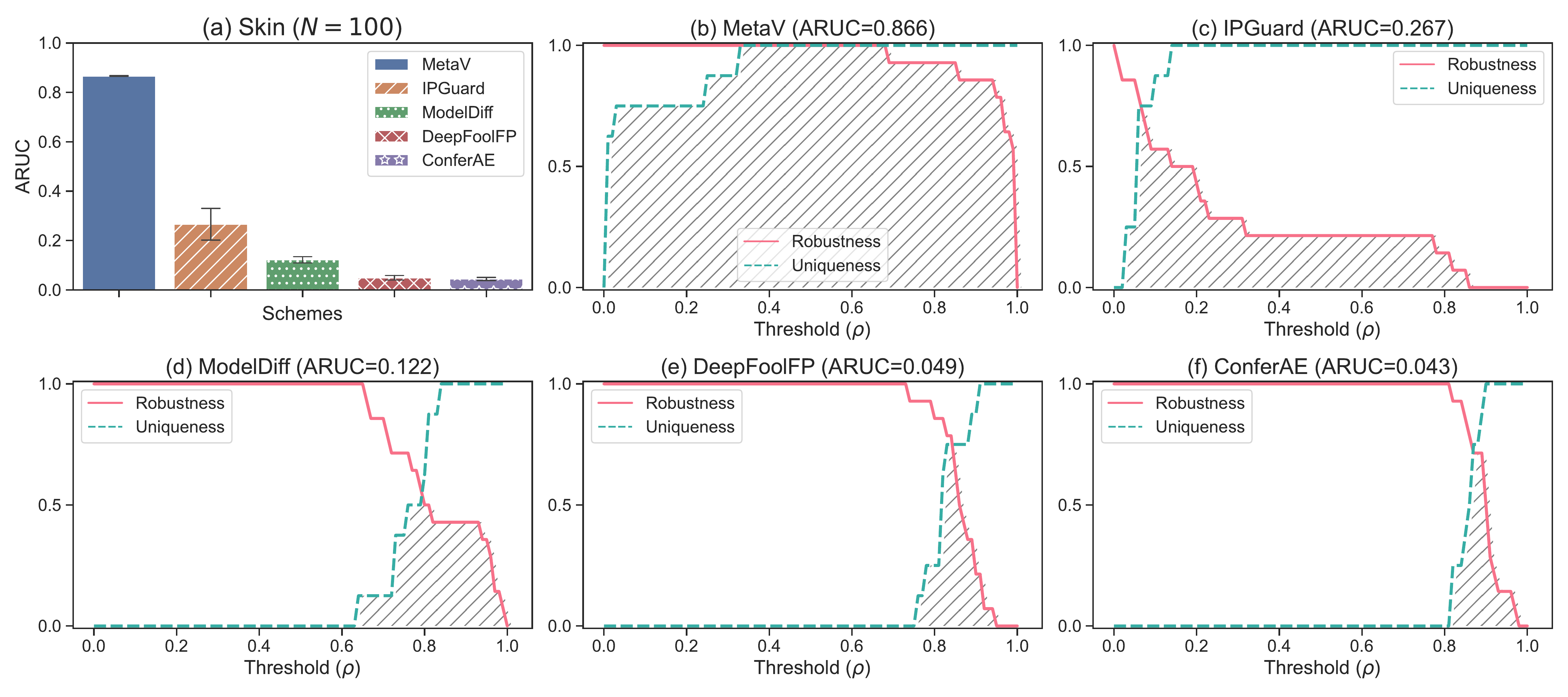}
\caption{(a) The ARUC of MetaV (\textit{Ours}) and baselines on Skin. (b)-(f): Curves of robustness and uniqueness of MetaV on Warfarin and Fashion, where the ARUC is reported in the figure title. }
\label{fig:skin_ru_suite_all}
\end{center}
\end{figure*}

\subsection{Baselines} We cover $4$  state-of-the-art fingerprinting schemes as the baselines for evaluating the effectiveness of MetaV under the classification settings. The baselines are respectively: 
\begin{itemize}
    \item \textit{IPGuard} \cite{Cao2019IPGuardPT}: IPGuard is one of the earliest fingerprinting schemes on classification models, which searches for a set of adversarial examples with a specified label near the decision boundary of the target model as the fingerprinting examples.  
    \item \textit{ConferAE} \cite{Lukas2019DeepNN}: ConferAE improves the design of IPGuard by further covering the distillation-based obfuscation techniques. Specifically, ConferAE constructs a set of adversarial examples on a prepared ensemble of positive/negative suspect models which may be implemented with different architecture from the target model. The generated adversarial examples are additionally required to be transferable from the target model to the positive suspect models but not to the negative ones. 
    \item \textit{DeepFoolFP} \cite{Wang2021FingerprintingDN}: DeepFoolFP literally leverages the DeepFool algorithm \cite{MoosaviDezfooli2016DeepFoolAS} to generate adversarial examples as the model fingerprints, the motivation of which is to improve the efficiency of fingerprint construction. For verification, the above model fingerprinting schemes define the matching rate as the ratio of the fingerprinting examples which are correctly classified by the suspect model into the specified class. 
    \item \textit{ModelDiff} \cite{Li2021ModelDiffTD}: ModelDiff is a very recent technique which is originally proposed to quantify the behavioral similarity between a pair of models. Specifically, the similarity is measured as the cosine similarity of two models' \textit{decision distance vector}, each element of which is the distance of the logits between a clean test input and an adversarial example derived from the input. A suspect model is verified if its behavioral similarity with the target model is smaller than a threshold. 
\end{itemize}
With no further specifications, the number of fingerprint examples, i.e., $N$, is set as $100$ for MetaV and the baselines by default. More details on the baselines are in Section \ref{sec:related}.

\subsection{Performance Metrics} Given a predefined threshold $\epsilon \in (0, 1)$, MetaV and all the baseline methods recognize a suspect model as positive when the matching rate of fingerprint verification is higher than $\rho$, or otherwise recognize the model as negative. In the evaluation, we following the evaluation protocol in \cite{Cao2019IPGuardPT} which is composed of the metrics below: 
\begin{enumerate}
\item \textbf{Robustness}/\textbf{Uniqueness} ($R(\rho)$/ $U(\rho)$): The robustness metric measures the proportion of positive suspect models also recognized as positive by the fingerprinting scheme, i.e., \textit{true positives}. 
\item \textbf{Uniqueness ($U(\rho)$)}: The uniqueness metric measures the proportion of negative suspect models also recognized as negative by the fingerprinting scheme, i.e.,\textit{true negatives}. 
\item \textbf{Area under the Robustness-Uniqueness Curves} (ARUC): ARUC measures the area of the intersecion region under the robustness and uniqueness when the threshold varies in $(0, 1)$, i.e., $\int_{0}^{1}\min\{R(\rho), U(\rho)\}d\rho$. A higher ARUC implies a more wider value range for the threshold to choose from to obtain simultaneously high robustness and uniqueness. ARUC is empirically calculated as the average $\min\{R(\rho), U(\rho)\}$ on $\{0, 1/L, \hdots, (L-1)/L, 1\}$ with $L = 100$. For all the experiments, we run $5$ repetitive experiments and report the average metric with the $95\%$ confidence interval. 
\end{enumerate}

\subsection{Other Implementation Details}
\subsubsection{Hyperparameter Setups} With no further specifications, we always set the number of fingerprint examples, i.e., $N$, for MetaV and the baselines as $100$ for fair comparisons. We set the learning rate in Algorithm 1 as $0.001$ and the number of iteration as $1000$. In all the three scenarios, we implement the meta-verifier $\mathcal{V}$ as a three-layer fully-connected neural network with the ReLU hidden layer size of $100$.

\subsubsection{Experimental Environment} All the defenses and experiments are implemented with PyTorch \cite{Paszke2019PyTorchAI}, an open-source software framework for numeric computation and deep learning. All our experiments are conducted on a Linux server running Ubuntu 16.04, one AMD Ryzen Threadripper 2990WX 32-core processor and 2 NVIDIA GTX RTX2080 GPUs.

\section{Results \& Analysis}
\label{sec:results}

\subsection{Comparison with Baselines} 
\begin{figure}[ht]
\begin{center}
\includegraphics[width=0.45\textwidth]{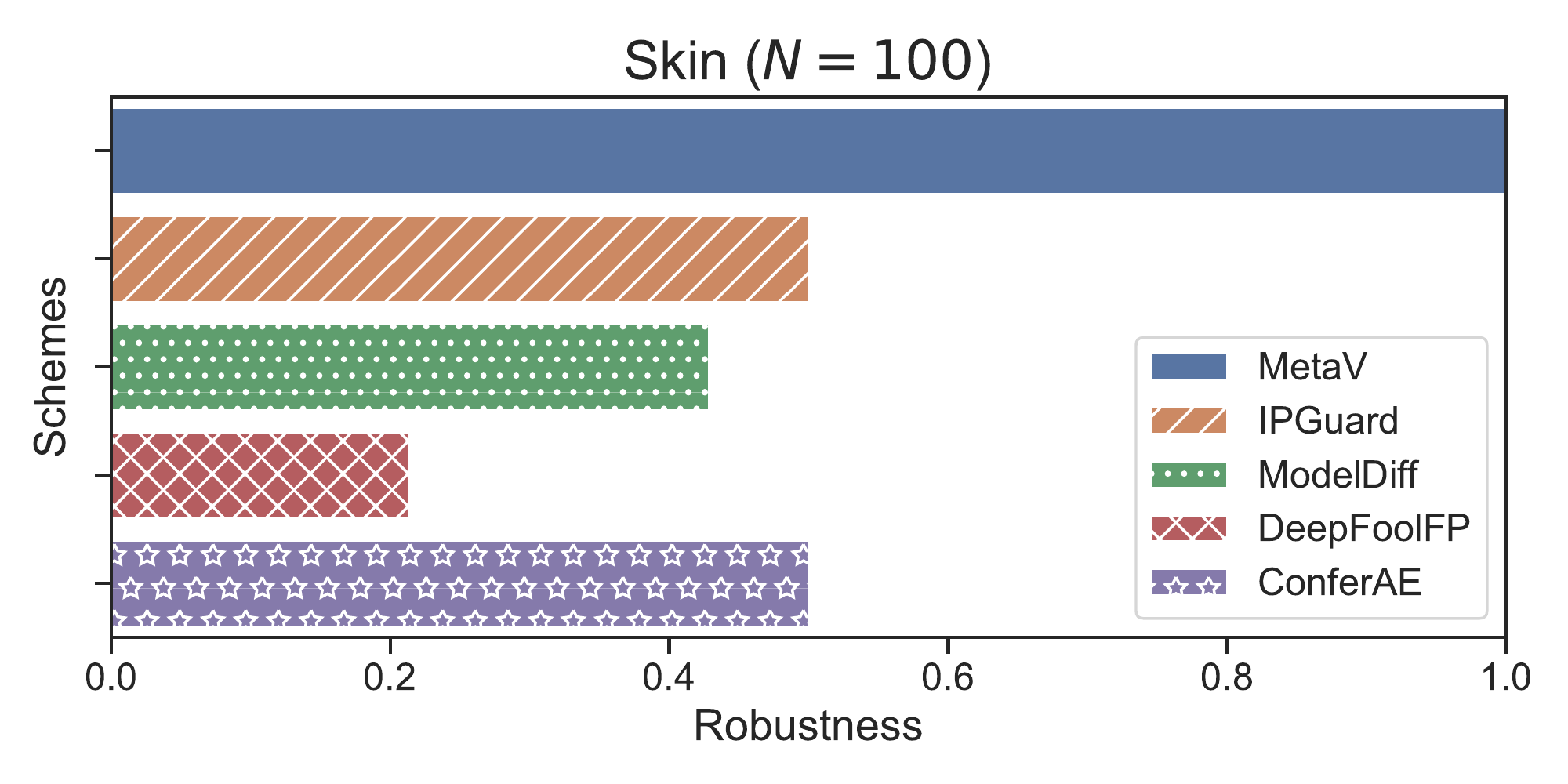}
\caption{The robustness when the threshold is set such that the uniqueness reaches $100\%$.}
\label{fig:skin_tps}
\end{center}
\end{figure}

First, we compare the performance of MetaV with $4$ state-of-the-art model fingerprinting schemes specifically designed for classifiers. Fig. \ref{fig:skin_tps} reports the robustness (i.e., true positives) when the threshold $\rho$ is set to allow the uniqueness (i.e., true negatives) to reach $100\%$ on the test set, along with the corresponding ARUC presented in Fig.\ref{fig:skin_ru_suite_all}(a).   
As Fig. \ref{fig:skin_tps} shows, our proposed MetaV is the only method which simultaneously achieves $100\%$ robustness and uniqueness in fingerprinting a stolen and adversarially obfuscated ResNet-18 classifier for skin cancer diagnosis. Besides, as we can see from Fig.\ref{fig:skin_ru_suite_all}(a), MetaV constructs the model fingerprint with the highest ARUC metric, i.e., $0.86\pm{0.01}$, among all the tested fingerprinting schemes, which improves the optimal baseline IPGuard by $0.59$ absolutely, i.e., a roughly $220\%$ relative improvement. As we construct a more diverse benchmark of suspect models compared with previous works, the ARUC of IPGuard is not as high as the results reported in \citet{Cao2019IPGuardPT}. Fig. \ref{fig:skin_ru_suite_all}(b)-(f) show the robustness and uniqueness curves of each fingerprint schemes.

\subsection{Time Efficiency of MetaV} 
\begin{figure}[ht]
\begin{center}
\includegraphics[width=0.45\textwidth]{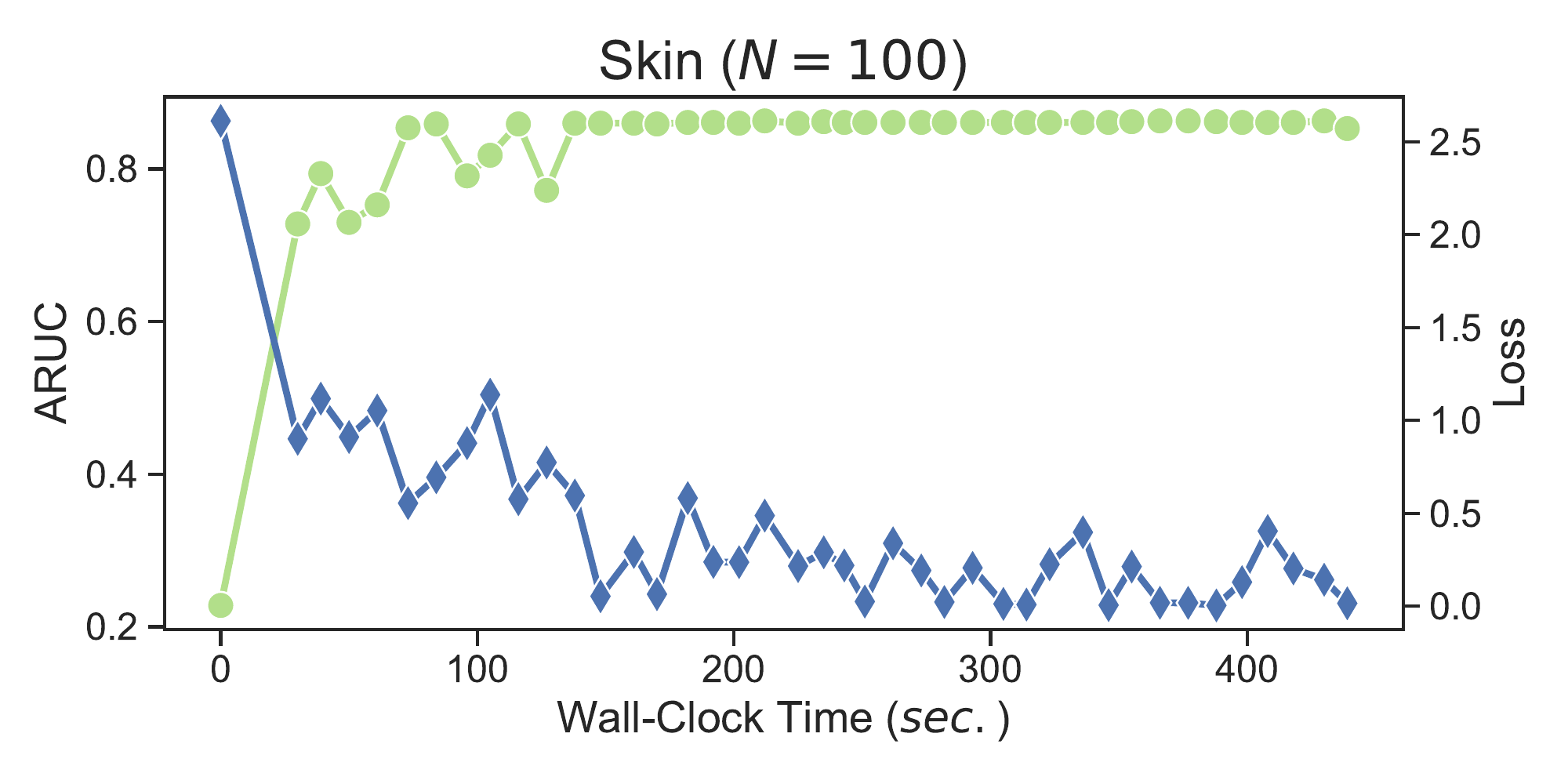}
\caption{The learning curves of MetaV for constructing a fingerprinting pair of ResNet-18 ($N=100$), where the x-axis shows the wall-clock time. }

\label{fig:trace}
\end{center}
\end{figure}

Next, we empirically study the learning behaviors and the time complexity of MetaV when constructing the fingerprint of a ResNet-18. As Fig. \ref{fig:trace} shows, the ARUC and loss curves demonstrate the time efficiency of MetaV in fingerprint construction. 

\begin{figure*}[ht]
\begin{center}
\includegraphics[width=1.0\textwidth]{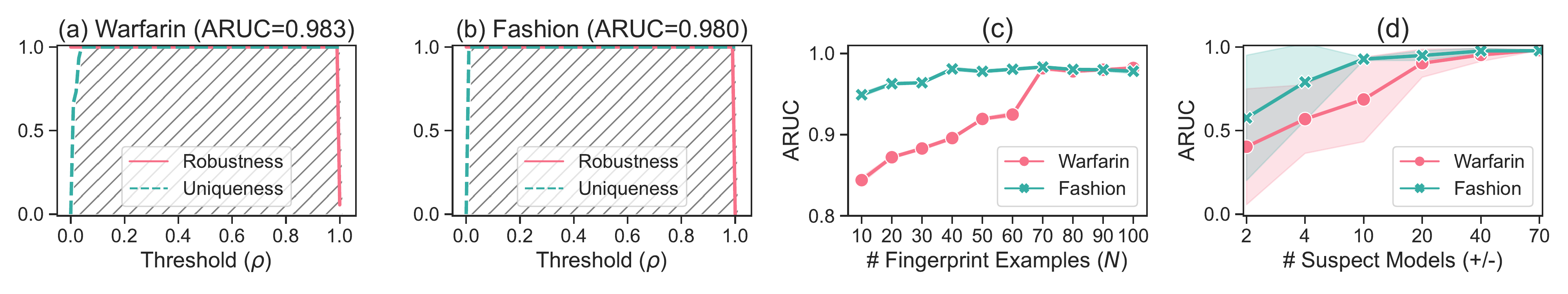}
\caption{(a)-(b): Curves of robustness and uniqueness of MetaV on Warfarin and Fashion, with the ARUC reported in the figure title. The curves of ARUC (c) when the number of fingerprint examples increases and (d) when the model ensemble is enlarged on Warfarin and Fashion.}
\label{fig:hyper_suite_all}
\end{center}
\end{figure*}
In less than $200$ seconds, MetaV stably constructs a fingerprinting pair which achieves an ARUC over $0.8$. Besides, Table \ref{tab:time} presents a tentative comparison on the time cost of each fingerprint scheme for constructing the corresponding model fingerprint to achieve the reported ARUC in Fig. \ref{fig:skin_ru_suite_all} and for fingerprint verification in the same experimental environment detailed in the experimental setting part. As is shown, MetaV is similarly efficient compared with the state-of-the-art fingerprinting schemes.  

\begin{table}[htbp]
  \centering
  \caption{Comparison of the time costs for fingerprint construction and verification (\textit{sec.}).}
    \begin{tabular}{lcc}
    \toprule
          & \textbf{Construction} & \textbf{Verification} \\
    \midrule
    \textbf{MetaV} & 202+11 & 7.2+0.8 \\
    \textbf{IPGuard} & 177+1 & 9.1+0.2 \\
    \textbf{ModelDiff} & 123+1 & 13.6+0.5 \\
    \textbf{DeepFoolFP} & 174+4 & 9.1+0.6 \\
    \textbf{ConferAE} & >2860 & 10.0+0.6 \\
    \bottomrule
    \end{tabular}%
  \label{tab:time}%
\end{table}%


\subsection{MetaV for Task-Independent Fingerprinting} Besides the substantial improvements in fingerprinting classifiers, more importantly, MetaV presents the first task-agnostic fingerprinting scheme which can be applied to more general application scenarios. To validate, we apply MetaV to fingerprint an MLP for regression (i.e., the \textbf{Warfarin} case) and a DCGAN for generative modeling (i.e., the \textbf{Fashion} case), which requires no modification on Algorithm \ref{alg:utaf_construction}, as MetaV is by design independent from either the internals or the functions of the target model.

Fig. \ref{fig:hyper_suite_all}(a)-(b) plot the curves of robustness and uniqueness of MetaV on Warfarin and Fashion when the threshold $\rho$ increases from $0$ to $1$, where the area of the shaded region is by definition the ARUC. As we can see, the robustness and the uniqueness remain $1$ unless the threshold is very close to $0$ or $1$. This results in an over $0.98$ ARUC for both the two scenarios which existing fingerprinting schemes can hardly handle.

\subsection{Number of Fingerprint Examples} 
\begin{figure}[ht]
\begin{center}
\includegraphics[width=0.45\textwidth]{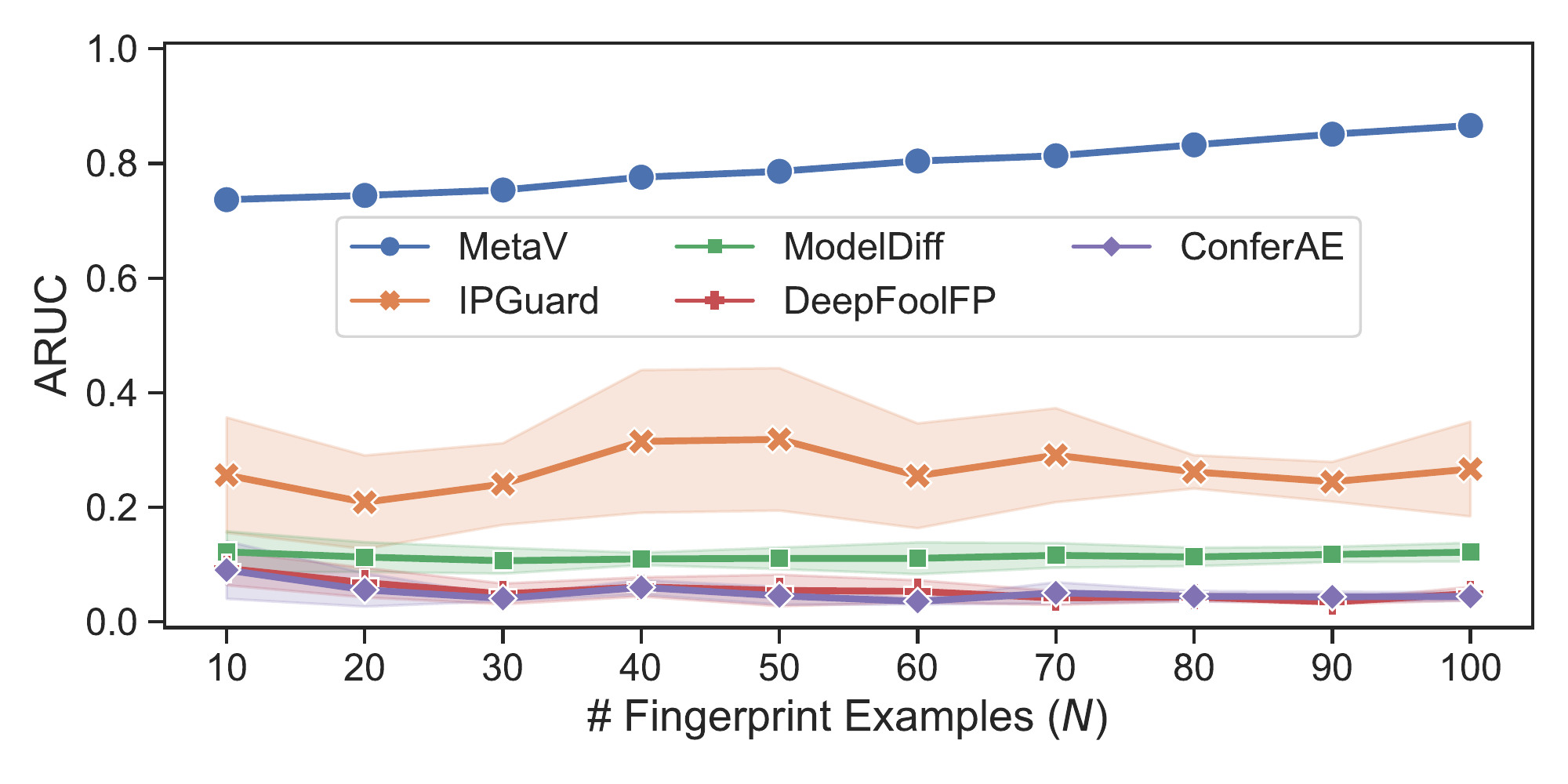}
\caption{The curves of ARUC when the number of fingerprint examples increases on Skin.}
\label{fig:aruc_fps_suite}
\end{center}
\end{figure}

We further study the influence of the number of fingerprint examples on the performance of MetaV and the baselines. Fig. \ref{fig:aruc_fps_suite}\&\ref{fig:hyper_suite_all}(c) presents the ARUC curves when the number of fingerprint examples, i.e., $N$, increases, on the classification and non-classification tasks respectively. As is shown, in all the three scenarios, the performance of MetaV increases stably when $N$ increases from $10$ to $100$. For example, when fingerprinting ResNet-18 on Skin, the ARUC of MetaV is about $1.2\times$ when $N$ is enlarged from $10$ to $100$. This is a desirable feature of MetaV as one would naturally expect a more accurate fingerprinting when more computing power is devoted to the construction of the model fingerprints. In comparison, the upward trend is unclear for all the baseline schemes. Similarly, enhanced performance is also observed on Warfarin and Fashion by about $3.2\%$ and $17.0\%$ respectively, which is noticeable considering the already high ARUC of MetaV when the number of fingerprint examples is $10$.

\subsection{Size of Prepared Model Ensemble} 
\begin{figure}[ht]
\begin{center}
\includegraphics[width=0.45\textwidth]{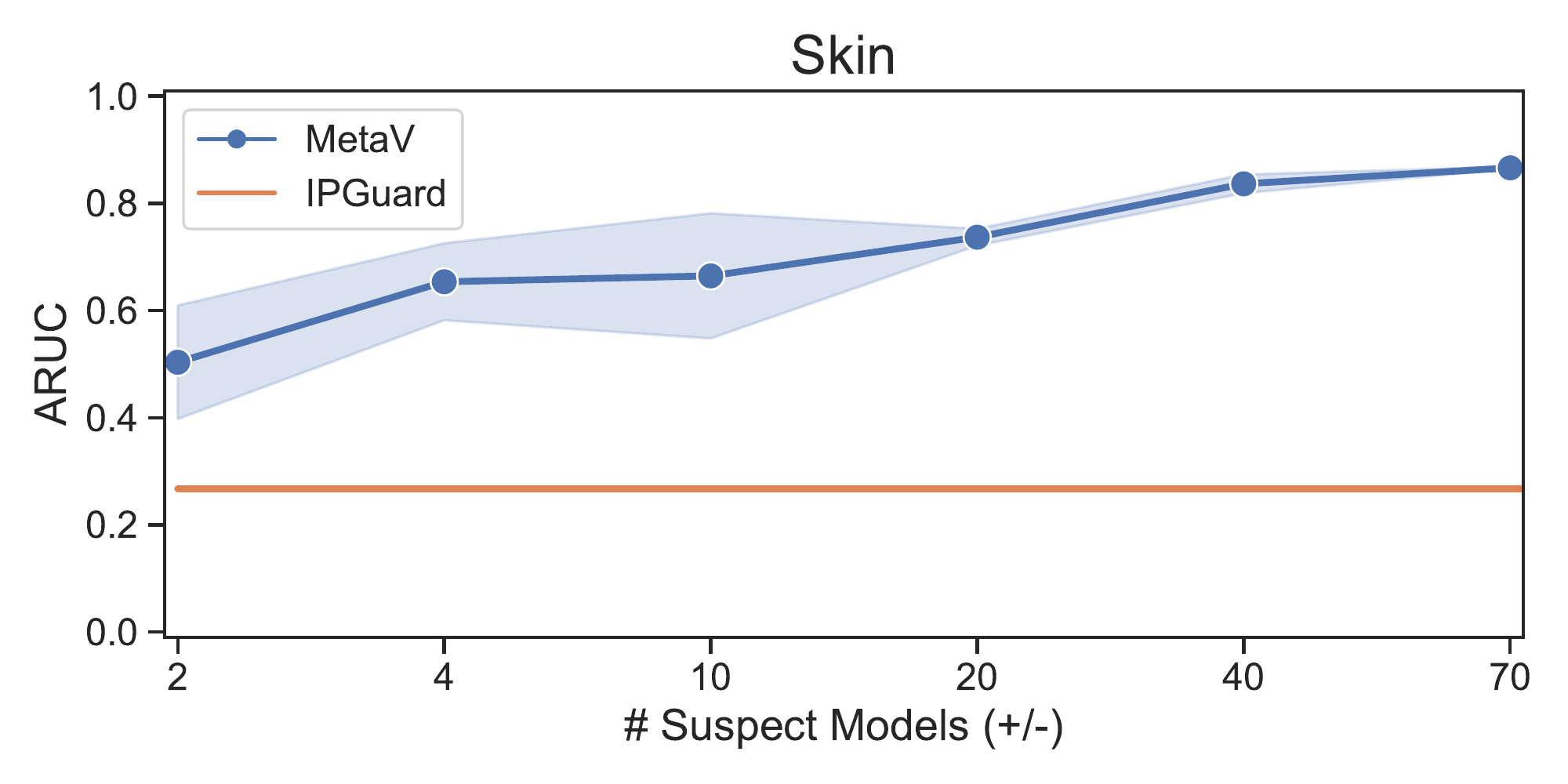}
\vspace{-0.2in}
\caption{The curves of ARUC when the model ensemble is enlarged on Skin.}
\label{fig:ensemble_aruc_skin}
\end{center}
\vspace{-0.2in}
\end{figure}

Finally, we provide quantitative results to analyze the impact of the model ensemble size on the performance of MetaV. We fix the number of fingerprint examples as $100$ and randomly sample different ratios of positive/negative suspect models from the full model ensemble for training MetaV. Fig. \ref{fig:ensemble_aruc_skin}\&\ref{fig:hyper_suite_all}(d) shows the ARUC curves on the three scenarios when the model ensemble size varies. As is shown, the ARUC of MetaV shows a steady upward trend when the model ensemble is enlarged, which conforms to our design principle that a more diverse set of crafted suspect models would help construct more unique and robust model fingerprints.  
\section{Conclusion}
In this paper, we present MetaV, the first task-agnostic model fingerprinting framework which: (a) substantially improves existing fingerprinting schemes on classification models in terms of fingerprint robustness and uniqueness, and, (b) more importantly, advances the capability of model piracy forensics to more general application scenarios. As it is a common challenge for \textit{any} novel fingerprinting methods to be evaluated on large-scale datasets (e.g., for evalaution on ImageNet, one has to train over $100$ suspect models on the dataset to construct the benchmark, which would 
incur over 50 days of computation on medium-end devices), we design our evaluation at the same scale of all our previous works by involving $32\times{32}$ images only. It would be meaningful for future works to cooperate with the industry to evaluate MetaV on larger datasets. Meanwhile, although MetaV by design has no assumptions on the input, the architecture, or the output of the target model, considering the impossibility of exhausting all possible task types, we mainly choose the three representative scenarios in our paper for evaluation. Future works may consider deploy and evaluate the effectiveness of MetaV on other typical learning tasks such as feature extraction, information retrieval and ranking.

 \bibliographystyle{ACM-Reference-Format}
 \bibliography{ref}


\begin{thebibliography}{46}


\ifx \showCODEN    \undefined \def \showCODEN     #1{\unskip}     \fi
\ifx \showDOI      \undefined \def \showDOI       #1{#1}\fi
\ifx \showISBNx    \undefined \def \showISBNx     #1{\unskip}     \fi
\ifx \showISBNxiii \undefined \def \showISBNxiii  #1{\unskip}     \fi
\ifx \showISSN     \undefined \def \showISSN      #1{\unskip}     \fi
\ifx \showLCCN     \undefined \def \showLCCN      #1{\unskip}     \fi
\ifx \shownote     \undefined \def \shownote      #1{#1}          \fi
\ifx \showarticletitle \undefined \def \showarticletitle #1{#1}   \fi
\ifx \showURL      \undefined \def \showURL       {\relax}        \fi
\providecommand\bibfield[2]{#2}
\providecommand\bibinfo[2]{#2}
\providecommand\natexlab[1]{#1}
\providecommand\showeprint[2][]{arXiv:#2}

\bibitem[\protect\citeauthoryear{??}{pyt}{[n.\,d.]}]%
        {pytorch_hub}
 \bibinfo{year}{[n.\,d.]}\natexlab{}.
\newblock \bibinfo{title}{PyTorch Hub}.
\newblock \bibinfo{howpublished}{https://pytorch.org/hub/}.
\newblock
\newblock
\shownote{Accessed: 2021-02-01}.


\bibitem[\protect\citeauthoryear{Adi, Baum, et~al\mbox{.}}{Adi
  et~al\mbox{.}}{2018}]%
        {Adi2018TurningYW}
\bibfield{author}{\bibinfo{person}{Y. Adi}, \bibinfo{person}{Carsten Baum},
  {et~al\mbox{.}}} \bibinfo{year}{2018}\natexlab{}.
\newblock \showarticletitle{Turning Your Weakness Into a Strength: Watermarking
  Deep Neural Networks by Backdooring}. In \bibinfo{booktitle}{\emph{USENIX
  Security Symposium}}.
\newblock


\bibitem[\protect\citeauthoryear{Aguinaldo, Chiang, Gain, Patil, Pearson, and
  Feizi}{Aguinaldo et~al\mbox{.}}{2019}]%
        {Aguinaldo2019CompressingGU}
\bibfield{author}{\bibinfo{person}{Angela Aguinaldo}, \bibinfo{person}{Ping-Yeh
  Chiang}, \bibinfo{person}{Alex Gain}, \bibinfo{person}{Ameya~D. Patil},
  \bibinfo{person}{Kolten Pearson}, {and} \bibinfo{person}{S. Feizi}.}
  \bibinfo{year}{2019}\natexlab{}.
\newblock \showarticletitle{Compressing GANs using Knowledge Distillation}.
\newblock \bibinfo{journal}{\emph{ArXiv}}  \bibinfo{volume}{abs/1902.00159}
  (\bibinfo{year}{2019}).
\newblock


\bibitem[\protect\citeauthoryear{Boenisch}{Boenisch}{2020}]%
        {Boenisch2020ASO}
\bibfield{author}{\bibinfo{person}{Franziska Boenisch}.}
  \bibinfo{year}{2020}\natexlab{}.
\newblock \showarticletitle{A Survey on Model Watermarking Neural Networks}.
\newblock \bibinfo{journal}{\emph{ArXiv}} (\bibinfo{year}{2020}).
\newblock


\bibitem[\protect\citeauthoryear{Cao, Jia, et~al\mbox{.}}{Cao
  et~al\mbox{.}}{2021}]%
        {Cao2019IPGuardPT}
\bibfield{author}{\bibinfo{person}{Xiaoyu Cao}, \bibinfo{person}{J. Jia},
  {et~al\mbox{.}}} \bibinfo{year}{2021}\natexlab{}.
\newblock \showarticletitle{IPGuard: Protecting the Intellectual Property of
  Deep Neural Networks via Fingerprinting the Classification Boundary}.
\newblock \bibinfo{journal}{\emph{AsiaCCS}} (\bibinfo{year}{2021}).
\newblock


\bibitem[\protect\citeauthoryear{Cao, Xiao, et~al\mbox{.}}{Cao
  et~al\mbox{.}}{2019}]%
        {Cao2019AdversarialSA}
\bibfield{author}{\bibinfo{person}{Yulong Cao}, \bibinfo{person}{Chaowei Xiao},
  {et~al\mbox{.}}} \bibinfo{year}{2019}\natexlab{}.
\newblock \showarticletitle{Adversarial Sensor Attack on LiDAR-based Perception
  in Autonomous Driving}.
\newblock \bibinfo{journal}{\emph{CCS}} (\bibinfo{year}{2019}).
\newblock


\bibitem[\protect\citeauthoryear{Carlini and Wagner}{Carlini and
  Wagner}{2017}]%
        {Carlini2017TowardsET}
\bibfield{author}{\bibinfo{person}{Nicholas Carlini} {and}
  \bibinfo{person}{David~A. Wagner}.} \bibinfo{year}{2017}\natexlab{}.
\newblock \showarticletitle{Towards Evaluating the Robustness of Neural
  Networks}.
\newblock \bibinfo{journal}{\emph{IEEE Symposium on Security and Privacy}}
  (\bibinfo{year}{2017}).
\newblock


\bibitem[\protect\citeauthoryear{Choromańska, Henaff, Mathieu, Arous, and
  LeCun}{Choromańska et~al\mbox{.}}{2015}]%
        {Choromaska2015TheLS}
\bibfield{author}{\bibinfo{person}{A. Choromańska}, \bibinfo{person}{Mikael
  Henaff}, \bibinfo{person}{Micha{\"e}l Mathieu}, \bibinfo{person}{G.~B.
  Arous}, {and} \bibinfo{person}{Y. LeCun}.} \bibinfo{year}{2015}\natexlab{}.
\newblock \showarticletitle{The Loss Surfaces of Multilayer Networks}. In
  \bibinfo{booktitle}{\emph{AISTATS}}.
\newblock


\bibitem[\protect\citeauthoryear{Clark, Luong, Khandelwal, Manning, and
  Le}{Clark et~al\mbox{.}}{2019}]%
        {Clark2019BAMBM}
\bibfield{author}{\bibinfo{person}{Kevin Clark}, \bibinfo{person}{Minh-Thang
  Luong}, \bibinfo{person}{Urvashi Khandelwal}, \bibinfo{person}{Christopher~D.
  Manning}, {and} \bibinfo{person}{Quoc~V. Le}.}
  \bibinfo{year}{2019}\natexlab{}.
\newblock \showarticletitle{BAM! Born-Again Multi-Task Networks for Natural
  Language Understanding}. In \bibinfo{booktitle}{\emph{ACL}}.
\newblock


\bibitem[\protect\citeauthoryear{Devlin, Chang, et~al\mbox{.}}{Devlin
  et~al\mbox{.}}{2019}]%
        {Devlin2019BERTPO}
\bibfield{author}{\bibinfo{person}{J. Devlin}, \bibinfo{person}{Ming-Wei
  Chang}, {et~al\mbox{.}}} \bibinfo{year}{2019}\natexlab{}.
\newblock \showarticletitle{BERT: Pre-training of Deep Bidirectional
  Transformers for Language Understanding}. In
  \bibinfo{booktitle}{\emph{NAACL-HLT}}.
\newblock


\bibitem[\protect\citeauthoryear{Esteva, Kuprel, et~al\mbox{.}}{Esteva
  et~al\mbox{.}}{2017}]%
        {Esteva2017DermatologistlevelCO}
\bibfield{author}{\bibinfo{person}{Andre Esteva}, \bibinfo{person}{B. Kuprel},
  {et~al\mbox{.}}} \bibinfo{year}{2017}\natexlab{}.
\newblock \showarticletitle{Dermatologist-level classification of skin cancer
  with deep neural networks}.
\newblock \bibinfo{journal}{\emph{Nature}} (\bibinfo{year}{2017}).
\newblock


\bibitem[\protect\citeauthoryear{Gou, Yu, Maybank, and Tao}{Gou
  et~al\mbox{.}}{2021}]%
        {Gou2021KnowledgeDA}
\bibfield{author}{\bibinfo{person}{Jianping Gou}, \bibinfo{person}{B. Yu},
  \bibinfo{person}{S. Maybank}, {and} \bibinfo{person}{D. Tao}.}
  \bibinfo{year}{2021}\natexlab{}.
\newblock \showarticletitle{Knowledge Distillation: A Survey}.
\newblock \bibinfo{journal}{\emph{Int. J. Comput. Vis.}}
  (\bibinfo{year}{2021}).
\newblock


\bibitem[\protect\citeauthoryear{Han, Pool, et~al\mbox{.}}{Han
  et~al\mbox{.}}{2015}]%
        {Han2015LearningBW}
\bibfield{author}{\bibinfo{person}{Song Han}, \bibinfo{person}{Jeff Pool},
  {et~al\mbox{.}}} \bibinfo{year}{2015}\natexlab{}.
\newblock \showarticletitle{Learning both Weights and Connections for Efficient
  Neural Network}.
\newblock \bibinfo{journal}{\emph{ArXiv}} (\bibinfo{year}{2015}).
\newblock


\bibitem[\protect\citeauthoryear{He, Zhang, et~al\mbox{.}}{He
  et~al\mbox{.}}{2016}]%
        {He2016DeepRL}
\bibfield{author}{\bibinfo{person}{Kaiming He}, \bibinfo{person}{X. Zhang},
  {et~al\mbox{.}}} \bibinfo{year}{2016}\natexlab{}.
\newblock \showarticletitle{Deep Residual Learning for Image Recognition}.
\newblock \bibinfo{journal}{\emph{CVPR}} (\bibinfo{year}{2016}),
  \bibinfo{pages}{770--778}.
\newblock


\bibitem[\protect\citeauthoryear{Heaton, Polson, et~al\mbox{.}}{Heaton
  et~al\mbox{.}}{2016}]%
        {Heaton2016DeepLF}
\bibfield{author}{\bibinfo{person}{J.~B. Heaton}, \bibinfo{person}{Nicholas~G.
  Polson}, {et~al\mbox{.}}} \bibinfo{year}{2016}\natexlab{}.
\newblock \showarticletitle{Deep Learning for Finance: Deep Portfolios}.
\newblock \bibinfo{journal}{\emph{Econometric Modeling: Capital Markets -
  Portfolio Theory eJournal}} (\bibinfo{year}{2016}).
\newblock


\bibitem[\protect\citeauthoryear{Hinton, Vinyals, and Dean}{Hinton
  et~al\mbox{.}}{2015}]%
        {Hinton2015DistillingTK}
\bibfield{author}{\bibinfo{person}{Geoffrey~E. Hinton}, \bibinfo{person}{Oriol
  Vinyals}, {and} \bibinfo{person}{J. Dean}.} \bibinfo{year}{2015}\natexlab{}.
\newblock \showarticletitle{Distilling the Knowledge in a Neural Network}.
\newblock \bibinfo{journal}{\emph{ArXiv}} (\bibinfo{year}{2015}).
\newblock


\bibitem[\protect\citeauthoryear{Huang, Liu, and Weinberger}{Huang
  et~al\mbox{.}}{2017}]%
        {Huang2017DenselyCC}
\bibfield{author}{\bibinfo{person}{Gao Huang}, \bibinfo{person}{Zhuang Liu},
  {and} \bibinfo{person}{Kilian~Q. Weinberger}.}
  \bibinfo{year}{2017}\natexlab{}.
\newblock \showarticletitle{Densely Connected Convolutional Networks}.
\newblock \bibinfo{journal}{\emph{2017 IEEE Conference on Computer Vision and
  Pattern Recognition (CVPR)}} (\bibinfo{year}{2017}),
  \bibinfo{pages}{2261--2269}.
\newblock


\bibitem[\protect\citeauthoryear{Iandola, Moskewicz, Ashraf, Han, Dally, and
  Keutzer}{Iandola et~al\mbox{.}}{2016}]%
        {Iandola2016SqueezeNetAA}
\bibfield{author}{\bibinfo{person}{Forrest~N. Iandola}, \bibinfo{person}{M.
  Moskewicz}, \bibinfo{person}{Khalid Ashraf}, \bibinfo{person}{Song Han},
  \bibinfo{person}{W. Dally}, {and} \bibinfo{person}{K. Keutzer}.}
  \bibinfo{year}{2016}\natexlab{}.
\newblock \showarticletitle{SqueezeNet: AlexNet-level accuracy with 50x fewer
  parameters and <1MB model size}.
\newblock \bibinfo{journal}{\emph{ArXiv}}  \bibinfo{volume}{abs/1602.07360}
  (\bibinfo{year}{2016}).
\newblock


\bibitem[\protect\citeauthoryear{Jeong, Ryu, and Hur}{Jeong
  et~al\mbox{.}}{2021}]%
        {Jeong2021NeuralNS}
\bibfield{author}{\bibinfo{person}{Hoyong Jeong}, \bibinfo{person}{Dohyun Ryu},
  {and} \bibinfo{person}{Junbeom Hur}.} \bibinfo{year}{2021}\natexlab{}.
\newblock \showarticletitle{Neural Network Stealing via Meltdown}.
\newblock \bibinfo{journal}{\emph{ICOIN}} (\bibinfo{year}{2021}),
  \bibinfo{pages}{36--38}.
\newblock


\bibitem[\protect\citeauthoryear{Juuti, Szyller, et~al\mbox{.}}{Juuti
  et~al\mbox{.}}{2019}]%
        {Juuti2019PRADAPA}
\bibfield{author}{\bibinfo{person}{Mika Juuti}, \bibinfo{person}{Sebastian
  Szyller}, {et~al\mbox{.}}} \bibinfo{year}{2019}\natexlab{}.
\newblock \showarticletitle{PRADA: Protecting Against DNN Model Stealing
  Attacks}.
\newblock \bibinfo{journal}{\emph{EuroS\&P}} (\bibinfo{year}{2019}).
\newblock


\bibitem[\protect\citeauthoryear{Kingma and Ba}{Kingma and Ba}{2015}]%
        {Kingma2015AdamAM}
\bibfield{author}{\bibinfo{person}{Diederik~P. Kingma} {and}
  \bibinfo{person}{Jimmy Ba}.} \bibinfo{year}{2015}\natexlab{}.
\newblock \showarticletitle{Adam: A Method for Stochastic Optimization}.
\newblock \bibinfo{journal}{\emph{CoRR}}  \bibinfo{volume}{abs/1412.6980}
  (\bibinfo{year}{2015}).
\newblock


\bibitem[\protect\citeauthoryear{Krizhevsky}{Krizhevsky}{2014}]%
        {Krizhevsky2014OneWT}
\bibfield{author}{\bibinfo{person}{A. Krizhevsky}.}
  \bibinfo{year}{2014}\natexlab{}.
\newblock \showarticletitle{One weird trick for parallelizing convolutional
  neural networks}.
\newblock \bibinfo{journal}{\emph{ArXiv}}  \bibinfo{volume}{abs/1404.5997}
  (\bibinfo{year}{2014}).
\newblock


\bibitem[\protect\citeauthoryear{Li, Kadav, et~al\mbox{.}}{Li
  et~al\mbox{.}}{2017}]%
        {Li2017PruningFF}
\bibfield{author}{\bibinfo{person}{Hao Li}, \bibinfo{person}{Asim Kadav},
  {et~al\mbox{.}}} \bibinfo{year}{2017}\natexlab{}.
\newblock \showarticletitle{Pruning Filters for Efficient ConvNets}.
\newblock \bibinfo{journal}{\emph{ArXiv}} (\bibinfo{year}{2017}).
\newblock


\bibitem[\protect\citeauthoryear{Li, Zhang, et~al\mbox{.}}{Li
  et~al\mbox{.}}{2021}]%
        {Li2021ModelDiffTD}
\bibfield{author}{\bibinfo{person}{Yuanchun Li}, \bibinfo{person}{Ziqi Zhang},
  {et~al\mbox{.}}} \bibinfo{year}{2021}\natexlab{}.
\newblock \showarticletitle{ModelDiff: testing-based DNN similarity comparison
  for model reuse detection}.
\newblock \bibinfo{journal}{\emph{ISSTA}} (\bibinfo{year}{2021}).
\newblock


\bibitem[\protect\citeauthoryear{Lukas, Zhang, et~al\mbox{.}}{Lukas
  et~al\mbox{.}}{2021}]%
        {Lukas2019DeepNN}
\bibfield{author}{\bibinfo{person}{Nils Lukas}, \bibinfo{person}{Yuxuan Zhang},
  {et~al\mbox{.}}} \bibinfo{year}{2021}\natexlab{}.
\newblock \showarticletitle{Deep Neural Network Fingerprinting by Conferrable
  Adversarial Examples}.
\newblock \bibinfo{journal}{\emph{ICLR}} (\bibinfo{year}{2021}).
\newblock


\bibitem[\protect\citeauthoryear{Moosavi-Dezfooli, Fawzi,
  et~al\mbox{.}}{Moosavi-Dezfooli et~al\mbox{.}}{2016}]%
        {MoosaviDezfooli2016DeepFoolAS}
\bibfield{author}{\bibinfo{person}{Seyed-Mohsen Moosavi-Dezfooli},
  \bibinfo{person}{Alhussein Fawzi}, {et~al\mbox{.}}}
  \bibinfo{year}{2016}\natexlab{}.
\newblock \showarticletitle{DeepFool: A Simple and Accurate Method to Fool Deep
  Neural Networks}.
\newblock \bibinfo{journal}{\emph{CVPR}} (\bibinfo{year}{2016}).
\newblock


\bibitem[\protect\citeauthoryear{Paszke, Gross, Massa, Lerer, Bradbury, Chanan,
  Killeen, Lin, Gimelshein, Antiga, Desmaison, K{\"o}pf, Yang, DeVito, Raison,
  Tejani, Chilamkurthy, Steiner, Fang, Bai, and Chintala}{Paszke
  et~al\mbox{.}}{2019}]%
        {Paszke2019PyTorchAI}
\bibfield{author}{\bibinfo{person}{Adam Paszke}, \bibinfo{person}{S. Gross},
  \bibinfo{person}{Francisco Massa}, \bibinfo{person}{Adam Lerer},
  \bibinfo{person}{James Bradbury}, \bibinfo{person}{Gregory Chanan},
  \bibinfo{person}{Trevor Killeen}, \bibinfo{person}{Zeming Lin},
  \bibinfo{person}{N. Gimelshein}, \bibinfo{person}{L. Antiga},
  \bibinfo{person}{Alban Desmaison}, \bibinfo{person}{Andreas K{\"o}pf},
  \bibinfo{person}{E. Yang}, \bibinfo{person}{Zach DeVito},
  \bibinfo{person}{Martin Raison}, \bibinfo{person}{Alykhan Tejani},
  \bibinfo{person}{Sasank Chilamkurthy}, \bibinfo{person}{Benoit Steiner},
  \bibinfo{person}{Lu Fang}, \bibinfo{person}{Junjie Bai}, {and}
  \bibinfo{person}{Soumith Chintala}.} \bibinfo{year}{2019}\natexlab{}.
\newblock \showarticletitle{PyTorch: An Imperative Style, High-Performance Deep
  Learning Library}. In \bibinfo{booktitle}{\emph{NeurIPS}}.
\newblock


\bibitem[\protect\citeauthoryear{Radford, Metz, and Chintala}{Radford
  et~al\mbox{.}}{2016}]%
        {Radford2016UnsupervisedRL}
\bibfield{author}{\bibinfo{person}{Alec Radford}, \bibinfo{person}{Luke Metz},
  {and} \bibinfo{person}{Soumith Chintala}.} \bibinfo{year}{2016}\natexlab{}.
\newblock \showarticletitle{Unsupervised Representation Learning with Deep
  Convolutional Generative Adversarial Networks}.
\newblock \bibinfo{journal}{\emph{CoRR}}  \bibinfo{volume}{abs/1511.06434}
  (\bibinfo{year}{2016}).
\newblock


\bibitem[\protect\citeauthoryear{Real, Aggarwal, Huang, and Le}{Real
  et~al\mbox{.}}{2019}]%
        {Real2019RegularizedEF}
\bibfield{author}{\bibinfo{person}{Esteban Real}, \bibinfo{person}{A.
  Aggarwal}, \bibinfo{person}{Y. Huang}, {and} \bibinfo{person}{Quoc~V. Le}.}
  \bibinfo{year}{2019}\natexlab{}.
\newblock \showarticletitle{Regularized Evolution for Image Classifier
  Architecture Search}. In \bibinfo{booktitle}{\emph{AAAI}}.
\newblock


\bibitem[\protect\citeauthoryear{Regazzoni, Palmieri, et~al\mbox{.}}{Regazzoni
  et~al\mbox{.}}{2021}]%
        {Regazzoni2021ProtectingAI}
\bibfield{author}{\bibinfo{person}{F. Regazzoni}, \bibinfo{person}{P.
  Palmieri}, {et~al\mbox{.}}} \bibinfo{year}{2021}\natexlab{}.
\newblock \showarticletitle{Protecting artificial intelligence IPs: a survey of
  watermarking and fingerprinting for machine learning}.
\newblock \bibinfo{journal}{\emph{CAAI Transactions on Intelligence
  Technology}} (\bibinfo{year}{2021}).
\newblock


\bibitem[\protect\citeauthoryear{Rouhani, Chen, et~al\mbox{.}}{Rouhani
  et~al\mbox{.}}{2018}]%
        {Rouhani2018DeepSignsAG}
\bibfield{author}{\bibinfo{person}{B. Rouhani}, \bibinfo{person}{Huili Chen},
  {et~al\mbox{.}}} \bibinfo{year}{2018}\natexlab{}.
\newblock \showarticletitle{DeepSigns: A Generic Watermarking Framework for IP
  Protection of Deep Learning Models}.
\newblock \bibinfo{journal}{\emph{ArXiv}} (\bibinfo{year}{2018}).
\newblock


\bibitem[\protect\citeauthoryear{Simonyan and Zisserman}{Simonyan and
  Zisserman}{2015}]%
        {Simonyan2015VeryDC}
\bibfield{author}{\bibinfo{person}{K. Simonyan} {and} \bibinfo{person}{Andrew
  Zisserman}.} \bibinfo{year}{2015}\natexlab{}.
\newblock \showarticletitle{Very Deep Convolutional Networks for Large-Scale
  Image Recognition}.
\newblock \bibinfo{journal}{\emph{ArXiv}} (\bibinfo{year}{2015}).
\newblock


\bibitem[\protect\citeauthoryear{Szegedy, Zaremba, Sutskever, Bruna,
  et~al\mbox{.}}{Szegedy et~al\mbox{.}}{2014}]%
        {Szegedy2014IntriguingPO}
\bibfield{author}{\bibinfo{person}{Christian Szegedy}, \bibinfo{person}{W.
  Zaremba}, \bibinfo{person}{Ilya Sutskever}, \bibinfo{person}{Joan Bruna},
  {et~al\mbox{.}}} \bibinfo{year}{2014}\natexlab{}.
\newblock \showarticletitle{Intriguing properties of neural networks}.
\newblock \bibinfo{journal}{\emph{ArXiv}} (\bibinfo{year}{2014}).
\newblock


\bibitem[\protect\citeauthoryear{Tram{\`e}r, Zhang, et~al\mbox{.}}{Tram{\`e}r
  et~al\mbox{.}}{2016}]%
        {Tramr2016StealingML}
\bibfield{author}{\bibinfo{person}{Florian Tram{\`e}r}, \bibinfo{person}{F.
  Zhang}, {et~al\mbox{.}}} \bibinfo{year}{2016}\natexlab{}.
\newblock \showarticletitle{Stealing Machine Learning Models via Prediction
  APIs}. In \bibinfo{booktitle}{\emph{USENIX Security}}.
\newblock


\bibitem[\protect\citeauthoryear{Truda and Marais}{Truda and Marais}{2019}]%
        {Truda2019WarfarinDE}
\bibfield{author}{\bibinfo{person}{G. Truda} {and} \bibinfo{person}{P.
  Marais}.} \bibinfo{year}{2019}\natexlab{}.
\newblock \showarticletitle{Warfarin dose estimation on multiple datasets with
  automated hyperparameter optimisation and a novel software framework}.
\newblock \bibinfo{journal}{\emph{ArXiv}} (\bibinfo{year}{2019}).
\newblock


\bibitem[\protect\citeauthoryear{Uchida, Nagai, et~al\mbox{.}}{Uchida
  et~al\mbox{.}}{2017}]%
        {Uchida2017EmbeddingWI}
\bibfield{author}{\bibinfo{person}{Y. Uchida}, \bibinfo{person}{Yuki Nagai},
  {et~al\mbox{.}}} \bibinfo{year}{2017}\natexlab{}.
\newblock \showarticletitle{Embedding Watermarks into Deep Neural Networks}.
\newblock \bibinfo{journal}{\emph{ICMR}} (\bibinfo{year}{2017}).
\newblock


\bibitem[\protect\citeauthoryear{Wang and Chang}{Wang and Chang}{2021}]%
        {Wang2021FingerprintingDN}
\bibfield{author}{\bibinfo{person}{Si Wang} {and} \bibinfo{person}{Chip-Hong
  Chang}.} \bibinfo{year}{2021}\natexlab{}.
\newblock \showarticletitle{Fingerprinting Deep Neural Networks - a DeepFool
  Approach}.
\newblock \bibinfo{journal}{\emph{ISCAS}} (\bibinfo{year}{2021}).
\newblock


\bibitem[\protect\citeauthoryear{Whirl-Carrillo, McDonagh,
  et~al\mbox{.}}{Whirl-Carrillo et~al\mbox{.}}{2012}]%
        {WhirlCarrillo2012PharmacogenomicsKF}
\bibfield{author}{\bibinfo{person}{M. Whirl-Carrillo}, \bibinfo{person}{E.
  McDonagh}, {et~al\mbox{.}}} \bibinfo{year}{2012}\natexlab{}.
\newblock \showarticletitle{Pharmacogenomics Knowledge for Personalized
  Medicine}.
\newblock \bibinfo{journal}{\emph{Clinical Pharmacology \& Therapeutics}}
  (\bibinfo{year}{2012}).
\newblock


\bibitem[\protect\citeauthoryear{Xiao, Rasul, et~al\mbox{.}}{Xiao
  et~al\mbox{.}}{2017}]%
        {Xiao2017FashionMNISTAN}
\bibfield{author}{\bibinfo{person}{H. Xiao}, \bibinfo{person}{K. Rasul},
  {et~al\mbox{.}}} \bibinfo{year}{2017}\natexlab{}.
\newblock \showarticletitle{Fashion-MNIST: a Novel Image Dataset for
  Benchmarking Machine Learning Algorithms}.
\newblock \bibinfo{journal}{\emph{ArXiv}} (\bibinfo{year}{2017}).
\newblock


\bibitem[\protect\citeauthoryear{Xu, Li, et~al\mbox{.}}{Xu
  et~al\mbox{.}}{2020}]%
        {Xu2020IdentityBF}
\bibfield{author}{\bibinfo{person}{Xiangrui Xu}, \bibinfo{person}{Y. Li},
  {et~al\mbox{.}}} \bibinfo{year}{2020}\natexlab{}.
\newblock \showarticletitle{“Identity Bracelets” for Deep Neural Networks}.
\newblock \bibinfo{journal}{\emph{IEEE Access}} (\bibinfo{year}{2020}).
\newblock


\bibitem[\protect\citeauthoryear{Yan, Fletcher, and Torrellas}{Yan
  et~al\mbox{.}}{2020}]%
        {Yan2020CacheTL}
\bibfield{author}{\bibinfo{person}{Mengjia Yan},
  \bibinfo{person}{Christopher~W. Fletcher}, {and} \bibinfo{person}{J.
  Torrellas}.} \bibinfo{year}{2020}\natexlab{}.
\newblock \showarticletitle{Cache Telepathy: Leveraging Shared Resource Attacks
  to Learn DNN Architectures}.
\newblock \bibinfo{journal}{\emph{USENIX Security}} (\bibinfo{year}{2020}).
\newblock


\bibitem[\protect\citeauthoryear{Yang, Shi, et~al\mbox{.}}{Yang
  et~al\mbox{.}}{2020}]%
        {Yang2020MedMNISTCD}
\bibfield{author}{\bibinfo{person}{Jiancheng Yang}, \bibinfo{person}{R. Shi},
  {et~al\mbox{.}}} \bibinfo{year}{2020}\natexlab{}.
\newblock \showarticletitle{MedMNIST Classification Decathlon: A Lightweight
  AutoML Benchmark for Medical Image Analysis}.
\newblock \bibinfo{journal}{\emph{ArXiv}} (\bibinfo{year}{2020}).
\newblock


\bibitem[\protect\citeauthoryear{Yu, Yang, Zhang, Tsai, Ho, and Jin}{Yu
  et~al\mbox{.}}{2020}]%
        {Yu2020CloudLeakLD}
\bibfield{author}{\bibinfo{person}{Honggang Yu}, \bibinfo{person}{Kaichen
  Yang}, \bibinfo{person}{Teng Zhang}, \bibinfo{person}{Yun-Yun Tsai},
  \bibinfo{person}{Tsung-Yi Ho}, {and} \bibinfo{person}{Yier Jin}.}
  \bibinfo{year}{2020}\natexlab{}.
\newblock \showarticletitle{CloudLeak: Large-Scale Deep Learning Models
  Stealing Through Adversarial Examples}. In \bibinfo{booktitle}{\emph{NDSS}}.
\newblock


\bibitem[\protect\citeauthoryear{Zhang, Gu, Jang, Wu, Stoecklin, Huang, and
  Molloy}{Zhang et~al\mbox{.}}{2018}]%
        {Zhang2018BackdoorWatermark}
\bibfield{author}{\bibinfo{person}{Jialong Zhang}, \bibinfo{person}{Zhongshu
  Gu}, \bibinfo{person}{Jiyong Jang}, \bibinfo{person}{Hui Wu},
  \bibinfo{person}{Marc~Ph Stoecklin}, \bibinfo{person}{Heqing Huang}, {and}
  \bibinfo{person}{Ian Molloy}.} \bibinfo{year}{2018}\natexlab{}.
\newblock \showarticletitle{Protecting intellectual property of deep neural
  networks with watermarking}.
\newblock \bibinfo{journal}{\emph{AsiaCCS}} (\bibinfo{year}{2018}).
\newblock


\bibitem[\protect\citeauthoryear{Zhao, Hu, et~al\mbox{.}}{Zhao
  et~al\mbox{.}}{2020}]%
        {Zhao2020AFAAF}
\bibfield{author}{\bibinfo{person}{Jingjing Zhao}, \bibinfo{person}{Qingyue
  Hu}, {et~al\mbox{.}}} \bibinfo{year}{2020}\natexlab{}.
\newblock \showarticletitle{AFA: Adversarial fingerprinting authentication for
  deep neural networks}.
\newblock \bibinfo{journal}{\emph{Comput. Commun.}} (\bibinfo{year}{2020}).
\newblock


\bibitem[\protect\citeauthoryear{Zhou, Zhang, Peng, Zhang, Li, Xiong, and
  Zhang}{Zhou et~al\mbox{.}}{2021}]%
        {Zhou2021InformerBE}
\bibfield{author}{\bibinfo{person}{Haoyi Zhou}, \bibinfo{person}{Shanghang
  Zhang}, \bibinfo{person}{Jieqi Peng}, \bibinfo{person}{Shuai Zhang},
  \bibinfo{person}{Jianxin Li}, \bibinfo{person}{Hui Xiong}, {and}
  \bibinfo{person}{Wancai Zhang}.} \bibinfo{year}{2021}\natexlab{}.
\newblock \showarticletitle{Informer: Beyond Efficient Transformer for Long
  Sequence Time-Series Forecasting}. In \bibinfo{booktitle}{\emph{AAAI}}.
\newblock


\end{thebibliography}
\end{document}